\documentclass{aa}

\usepackage[]{graphicx}  

\usepackage{natbib,twoopt}
\bibpunct{(}{)}{;}{a}{}{,} 

\usepackage[normalem]{ulem}  

\usepackage[varg]{txfonts}
\usepackage{color}

\def\pow#1#2{#1$\times$10$^{#2}$}

\def\micron{$\mu$m}
\def\kms{$\mathrm{km}\,\mathrm{s}^{-1}$}  
\def\psqcm{$\mathrm{cm}^{-2}$}
\def\pccm{$\mathrm{cm}^{-3}$}
\def\Tkin{$T_\mathrm{kin}$}   

\def\vlsr{$V_\mathrm{lsr}$}  
\def\Eup{$E_\mathrm{up}$}  
\def\nHH{$n_\mathrm{H_2}$}  
\def\h{$^\mathrm{h}$}  
\def\m{$^\mathrm{m}$}  
\def\Msun{$M_\sun$}  
\def\h{$^\mathrm{h}$}  
\def\m{$^\mathrm{m}$}  
\def\Jyperbeam{Jy\,beam$^{-1}$}  

\definecolor{orange}{rgb}{0.8,0.4,0.0}
\definecolor{darkblue}{rgb}{0.0,0.0,0.4}

\def\HH{H$_2$}
\def\HCOplus{HCO$^+$}

\def\HHCO{H$_2$CO} 
\def\HHthCO{H$_2$$^{13}$CO} 
\def\HthCN{H$^{13}$CN}

\def\thCO{$^{13}$CO}
\def\CstO{C$^{17}$O}
\def\CeiO{C$^{18}$O}
\def\SOtwo{SO$_2$}
\def\thfSOtwo{$^{34}$SO$_2$}

\def\CtfS{C$^{34}$S}
\def\CCH{C$_2$H}

\defcitealias{jacobsen2018}{J2018} 
\defcitealias{lee_ki2016}{K.~I.~Lee et al.~2016}
\defcitealias{lee_j-e2017}{J.-E.~Lee et al.~2017}

\definecolor{orange}{rgb}{0.8,0.4,0.0}
\definecolor{darkblue}{rgb}{0.0,0.0,0.6}
\definecolor{darkred}{rgb}{0.75,0.0,0.0}

\def\markchanges{yes}  
\def\marked{yes}
\def\unmarked{no}
\ifx\markchanges\marked 
	\newcommand{\removed}[1]{\textcolor{darkred}{[}\sout{#1}\textcolor{darkred}{]}}  
	
\fi
\ifx\markchanges\unmarked 
	\newcommand{\removed}[1]{}  
	
\fi

\begin{document}

\title{The ALMA-PILS survey: Gas dynamics in IRAS\,16293$-$2422 and the connection between its two protostars}
\titlerunning{PILS: Gas dynamics in IRAS\,16293 and the connection between its two protostars}

\author{
M.~H.~D.~van der Wiel\inst{\ref{astron},\ref{nbi-starplan}} \and 
S.~K.~Jacobsen\inst{\ref{nbi-starplan}}  \and 
J.~K.~J\o rgensen\inst{\ref{nbi-starplan}} \and
T.~L.~Bourke\inst{\ref{SKA-office}} \and
L.~E.~Kristensen\inst{\ref{nbi-starplan}} \and 
P.~Bjerkeli\inst{\ref{onsala},\ref{nbi-starplan}} \and 
N.~M.~Murillo\inst{\ref{leiden}} \and 
H.~Calcutt\inst{\ref{nbi-starplan}} \and
H.~S.~P.~M\"uller\inst{\ref{koln}} \and 
A.~Coutens\inst{\ref{bordeaux}} \and 
M.~N.~Drozdovskaya\inst{\ref{csh-bern}} \and 
C.~Favre\inst{\ref{inaf-arcetri}} \and 
S.~F.~Wampfler\inst{\ref{csh-bern}} 
}

\institute{
ASTRON, Netherlands Institute for Radio Astronomy, Oude Hoogeveensedijk 4, 7991 PD Dwingeloo, The Netherlands \\ email: \texttt{mhd@vanderwiel.org}
\label{astron} 
\and
Centre for Star and Planet Formation, Niels Bohr Institute \& Natural History Museum of Denmark, University of Copenhagen, \O ster Voldgade 5--7, 1350 \mbox{Copenhagen~K}, Denmark
\label{nbi-starplan}
\and
SKA Organisation, Jodrell Bank Observatory, Lower Withington, Macclesfield, Cheshire SK11 9DL, UK
\label{SKA-office}
\and 
Department of Space, Earth and Environment, Chalmers University of Technology, Onsala Space Observatory, 43992 Onsala, Sweden
\label{onsala}
\and 
Leiden Observatory, Leiden University, P.O.~Box 9513, 2300 RA, Leiden, The Netherlands
\label{leiden}
\and
I.~Physikalisches Institut, Universit\"at zu K\"oln, Z\"ulpicher Str.~77, 50937 K\"oln, Germany
\label{koln} 
\and 
Laboratoire d'Astrophysique de Bordeaux, Univ.~Bordeaux, CNRS, B18N, all\'ee Geoffroy Saint-Hilaire, 33615 Pessac, France
\label{bordeaux}
\and 
Center for Space and Habitability, Universit\"at Bern, Sidlerstrasse 5, 3012 Bern, Switzerland
\label{csh-bern}
\and
INAF-Osservatorio Astrofisico di Arcetri, Largo E. Fermi 5, I-50125, Florence, Italy
\label{inaf-arcetri}
}

\date{\today}

\abstract
{ 
The majority of stars form in binary or higher order systems. 
The evolution of each protostar in a multiple system may start at different times and may progress differently. 
The Class 0 protostellar system IRAS\,16293$-$2422 contains two protostars, ``A'' and ``B'', separated by $\sim$600~au and embedded in a single, $10^4$~au scale envelope. Their relative evolutionary stages have been debated. 
}
{
We aim to study the relation and interplay between the two protostars A and B at spatial scales of 60~au up to $\sim$$10^3$~au.    
}
{
We selected molecular gas line transitions of the species CO, \HHCO, HCN, CS, SiO, and \CCH\ from the ALMA-PILS spectral imaging survey (329--363~GHz) and used them as tracers of kinematics, density, and temperature in the IRAS\,16293$-$2422 system. The angular resolution of the PILS data set allows us to study these quantities at a resolution of 0.5\arcsec\ (60 au at the distance of the source). 
}
{
Line-of-sight velocity maps of both optically thick and optically thin molecular lines reveal: (i) new manifestations of previously known outflows emanating from protostar~A; (ii) a kinematically quiescent bridge of dust and gas spanning between the two protostars, with an inferred density between \pow{4}{4}\,\pccm\ and $\sim$\pow{3}{7}\,\pccm; and (iii) a separate, straight filament seemingly connected to protostar~B seen only in \CCH, with a flat kinematic signature. Signs of various outflows, all emanating from source~A, are evidence of high-density and warmer gas; none of them coincide spatially and kinematically with the bridge.  
}
{ 
We hypothesize that the bridge arc is a remnant of filamentary substructure in the protostellar envelope material from which protostellar sources A and B have formed. 
One particular morphological structure appears to be due to outflowing gas impacting the quiescent bridge material. The continuing lack of clear outflow signatures unambiguously associated to protostar~B and the vertically extended shape derived for its disk-like structure lead us to conclude that source~B may be in an earlier evolutionary stage than source~A. 
}

\keywords{ISM: individual objects: IRAS\,16293 -- stars: formation -- circumstellar matter -- ISM: jets and outflows }

\maketitle

\section{Introduction}
\label{sec:intro}

The majority of currently forming stars are part of multiple systems \citep{chen2013,duchene2013,tobin2016a}. 
Two main scenarios have been proposed to form stellar systems of binary and higher order: disk fragmentation \citep{adams1989,bonnell1994a}, leading to close binaries (up to a few hundred au separation), and turbulent fragmentation of a natal protostellar envelope \citep{offner2010,pineda2015}, leading to wide separation binaries ($\gtrsim$1000~au). The difference in initial separation is often obfuscated at later stages by the effects of tidal interactions after the initial formation of protostars. Another observable difference that can help to distinguish between formation scenarios is that binaries formed  through disk instability may be more prone to exhibit aligned rotation axes, whereas rotation axes of those formed from turbulent fragmentation should be randomly distributed. The rotation axis of a protostar can be inferred through studying the morphology and kinematics of its disk-outflow system \citep{lee_ki2016}. 
High angular resolution observations of embedded protostellar systems suggest that both mechanisms of multiple star formation occur in nature (e.g., \citealt{tobin2016b}; \citealt{brinch2016}; \citealt{lee_j-e2017}). 
In both cases, remnant material of the natal body may manifest as a bridge between recently formed binary companions. The object of study in this paper is one of the most well-studied protostellar binary systems, IRAS\,16293$-$2422 (hereafter IRAS\,16293), which has such a bridge and exhibits kinematic signatures of active outflow and infall (see below). 

\object{IRAS\,16293} is a nearby, young, Class 0 protostellar system in the Ophiuchus cloud complex (see Sect.~2 of \citealt{jorgensen2016} for a review of the source). The distance to three other young stellar objects in the same cloud complex has been determined at 147$\pm$3~pc \citep{ortiz-leon2017a}, and trigonometric parallax measurements of water maser spots in IRAS\,16293 itself put it at a distance of 141$^{+30}_{-21}$~pc \citep{dzib2018}. 
However, to be consistent with earlier work and with detailed numerical modeling of the IRAS\,16293 system conducted by our team, we adopt a distance estimate of 120~pc based on extinction measurements and VLBI parallax measurements of two stars in the core of the complex \citep{loinard2008}. This value is within the 1$\sigma$ uncertainty margin of \citet{dzib2018}. In general, an 18\% increase in the source distance estimate (with an uncertainty margin of 15--20\% of its own) inflates projected linear scales by a factor 1.18, while quantities such as mass and luminosity, determined from broadband flux and column density, would scale by the distance squared, that is, a factor 1.39. 
With $d$=120~pc, the angular distance of 5.3\arcsec\ between the two submillimeter sources A and B corresponds to a projected distance of 636~au. The exact three-dimensional geometry of the two sources, their potential disks and the filament structure is unknown. Based on a total of 515 spectral lines from 54 molecular species detected in a Submillimeter Array (SMA) spectral survey, \citet{jorgensen2011} derived LSR velocities of sources~A and B of $+3.2$ and $+2.7$\,\kms, respectively (throughout this work, IRAS\,16293A will be abbreviated as ``source A'' and IRAS\,16293B as ``source B''). 
Different centroid velocities have been reported based on individual molecular lines: for source A, $+3.8$~\kms\ from \CstO, \CtfS\ \citep{favre2014a}, $+3.6$\,\kms\ from HCN \citep{takakuwa2007}; for source B, $+3.4$~\kms\ from CH$_3$OCHO, H$_2$CCO \citep{pineda2012}. 
Differences in the derived centroid velocity may stem from different angular resolution of the data sets, the different molecular tracers used, and/or varying optical depth of individual transitions of the same species, so that each traces a somewhat different ensemble of gas. 
All of the centroid velocities listed above fall well within the $\sim$3--4~\kms\ wide distributions of \vlsr\ values fitted by \citet{jorgensen2011}. 

\citet{caux2011} used single-dish observations to estimate the masses of the two protostellar sources. Assuming that the line broadening of the profiles is due to infalling motions, these authors arrive at $\sim$0.8~\Msun\ for source~A and 0.1~\Msun\ for source~B. 
The same mass of source~A was found by \citet{bottinelli2004c}, also assuming infalling motion to explain the observed line broadening, but using interferometric observations in which source~A is spatially separated from source~B. 
Alternatively, the line broadening could be caused by rotating motion instead of infall, as assumed for source~A by \citet{pineda2012}, who used observations of methyl formate (CH$_3$OCHO) and ketene (H$_2$CCO) with a 2.2\arcsec$\times$1.0\arcsec\ beam to infer Keplerian rotation around a central object of 0.53~\Msun. \citet{oya2016} found a mass varying between 0.5 and 1.0~\Msun, depending on the assumed inclination and centrifugal barrier radius. These studies show that the source masses are in the regime of low-mass protostars, but the various methods yield masses differing by up to a factor of two. 

The binary protostellar system IRAS\,16293 is embedded in an envelope with a radius of (6--8)$\times$$10^3$~au \citep[e.g.,][]{schoeier2002,crimier2010b}. Rotating, infalling motions of this envelope have been inferred from spectral line profiles \citep[e.g.,][]{menten1987,zhou1995,ceccarelli2000a,schoeier2002,takakuwa2007}. In a detailed velocity model posited by \citet{oya2016}, the large-scale envelope could transition into a Keplerian disk within the centrifugal barrier around source~A, as indicated by observations presented in \citet{favre2014a}. The latter study found rotation on 50--400~au scales, which could not be explained by simple Keplerian rotation around a point-mass, but needed to take into account the extra material of the enclosed mass at these scales. 

\begin{figure*}
	\resizebox{\hsize}{!}{\includegraphics{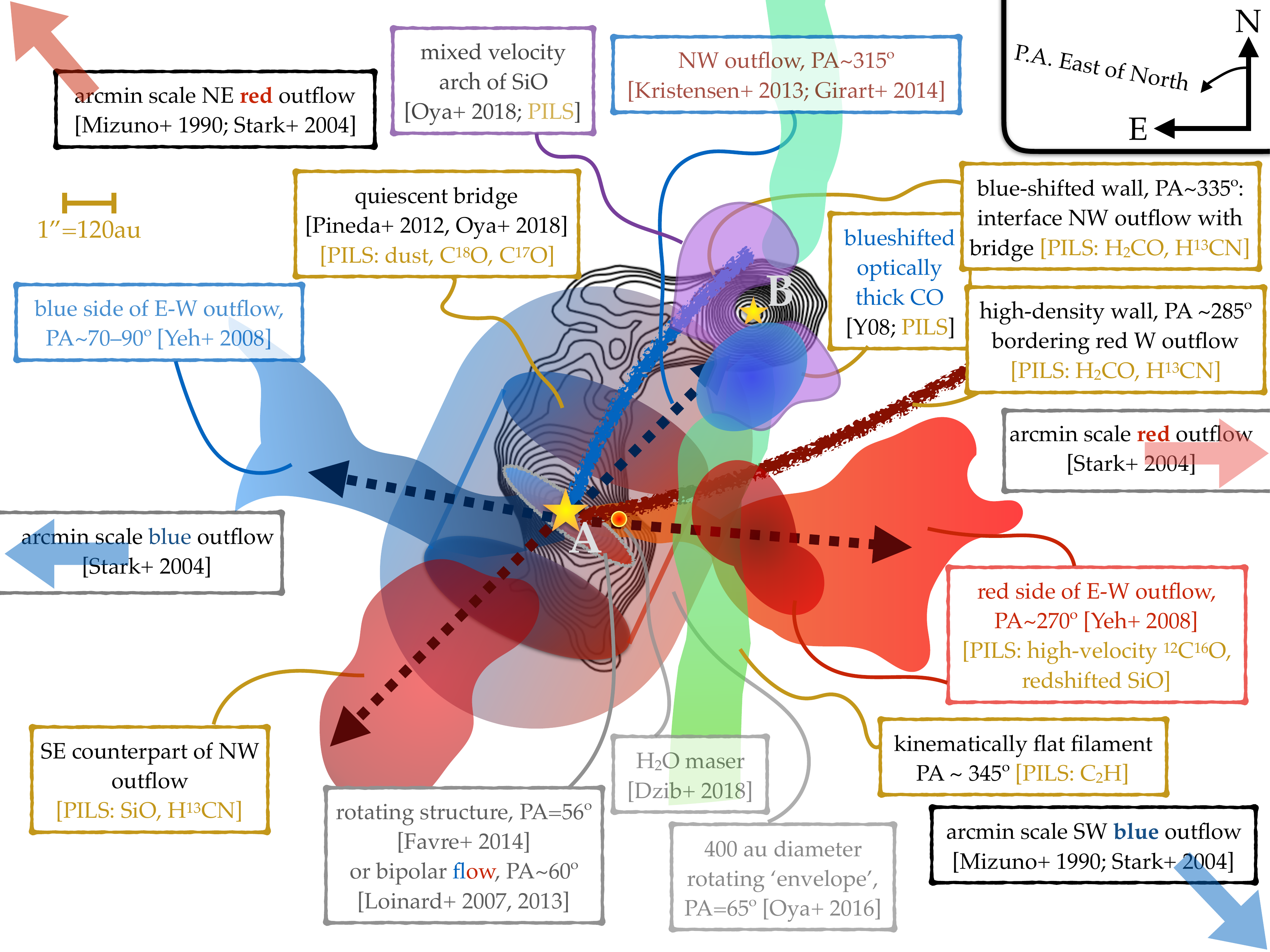}}
	\caption{Illustration of physical components surrounding and bridging protostars A and B in the IRAS\,16293 system, and outflows emanating from IRAS\,16293A. Thick, solid arrows at the edges of the panel point to scales beyond the $\sim$20\arcsec\ depicted in this illustration. Position angle (P.A.) is defined from north to east, as indicated in the top right. Each component is labeled by a rectangular box, with literature sources listed in square brackets. In all references given in square brackets in the illustration, `et al.' is abbreviated as `+': \citet{dzib2018}, \citet{girart2014},  \citet{kristensen2013a}, \citet{loinard2007b,loinard2013}, \citet{mizuno1990}, \citet{oya2016,oya2018}, \citet{stark2004}, \citet{yeh2008} [further abbreviated as `Y08' where needed]; `PILS' refers to structures observed in ALMA PILS observations (\citealt{jorgensen2016}, \citetalias{jacobsen2018}, and this work). 
	}
	\label{fig:outflowcartoon}
\end{figure*}

At least two major outflows have been observed from IRAS\,16293A: an east-west bipolar outflow \citep{yeh2008} and a northwest-southeast outflow pair \citep{kristensen2013a,girart2014}. On larger scales of $>$5000\,au, a northeast outflow has also been observed \citep{mizuno1990,stark2004}, the origin of which is speculated to be IRAS\,16293A, based on high-resolution CO images revealing a collimated structure near source~A (see Fig.~\ref{fig:outflowcartoon}). Combining single-dish observations of a broad range of HCO$^+$ spectral lines ($J$=1--0 up to $J$=13--12), \citet{quenard2018b} conclude that the spatially unresolved spectral line profiles are dominated by outflow contributions, explained by these authors using a model of the northwest-southeast outflow emanating from source~A. 
In contrast, any signs of outflows associated with IRAS\,16293B have long escaped detection. \citet{loinard2013} argued that the blueshifted emission found southeast of source~B is from a young monopolar outflow from source~B, but \citet{kristensen2013a} demonstrated, using the same data, that it could be a bow-shock from the northwest-southeast outflow from IRAS\,16293A. \citet{oya2018} presented SiO velocity maps indicative of a pole-on pair of outflows from source~B, but these authors also recognize that interaction with the outflow from source~A is not ruled out as a possible scenario.  

The quadruple outflow structure from source~A has led to speculations that source~A itself is a multiple system. Indeed, continuum observations at centimeter waves resolved IRAS\,16293A into two \citep{wootten1989,chandler2005} or even three components \citep{loinard2007b,pech2010}, with a separation of up to $\sim$0.5\arcsec. The 0.5\arcsec\ resolution maps from the ALMA PILS program appear to indicate a singularly peaked source at the position of source~A, both in continuum \citep{jorgensen2016} and in optically thin \CstO\ \citep{jacobsen2018}. However, the integrated \CstO~3--2 intensity is affected by contamination from more complex molecules in the dense, warm regions close to the two protostellar sources, which modifies the apparent morphology of emission integrated over channels bracketing the \CstO\ line frequency (see Fig.~\ref{fig:C17Ospectrum} in Appendix~\ref{sec:contamination}). Moreover, the 0.87~mm dust continuum is optically thick in the disk domains, which could hide any intrinsically multipeaked structure \citep{calcutt2018a}. In contrast, the PILS maps of various isotopologues of methyl cyanide in \citet{calcutt2018a}, for which the emission is integrated over a moving velocity interval to avoid incorporating contributions from other species, do show two clearly separated emission peaks. The methyl cyanide peak positions are consistent with those of radio continuum (3.6~cm) positions A1 and A2 from \citet{loinard2007b} after correction for proper motion \citep{pech2010}. The slight offset of $\sim$0.2\arcsec\ is likely due to pointing errors and uncertainties in the proper motion correction coefficients \citep{calcutt2018a}. 

Material spanning the interbinary region between IRAS\,16293A and B was first identified in millimeter observations from the BIMA array\footnote{BIMA stands for Berkeley-Illinois-Maryland Association, which operated and funded the nine-telescope array at Hat Creek in California until 2004, when its hardware was relocated and merged into the Combined Array for Research in Millimeter-wave Astronomy (CARMA).} by \citet{looney2000a}, who interpreted it as a circumbinary envelope. The same bridge material is also clearly detected in sensitive Atacama Large Millimeter/Submillimeter Array (ALMA) observations \citep{pineda2012,jorgensen2016}. Observations of polarized dust emission indicate that the magnetic field in the filamentary bridge is oriented along the long axis of the bridge \citep{rao2009,rao2014,sadavoy2018c}.
\citet{jacobsen2018} (hereafter \citetalias{jacobsen2018}) show that a dust filament model can match the observed spatial emission from high-resolution submillimeter continuum emission and \CstO~3--2 gas line emission. 
Similar structures observed in other multiple protostellar systems are interpreted as fragmentation of large-scale circum-multiple envelopes \citep{lee_j-e2017} or (Keplerian) rotating disks that have become locally gravitationally unstable \citep{tobin2016b,fernandez-lopez2017,dutrey2014,dutrey2016}. 

The exact evolutionary stage of the two protostars is unknown, partly because of the different inclination angles of sources~A (edge on; \citealt{pineda2012,oya2016}) and B (face on; \citealt{oya2018}). Given the outflow and infall signatures near source~A, its status as a protostar is firmly established. In contrast, the apparent quiescence of source~B has led to speculations that it is not a young protostar, but rather harbors a late-stage T~Tauri disk \citep{stark2004}. This interpretation was later called into question when spectral line signatures of infalling material were observed toward source~B \citep{chandler2005,pineda2012,jorgensen2012}. In conclusion, it is as yet unknown if the two protostars are at the same evolutionary stage, or that one may be more evolved than the other. 

The outflows from source~A, along with the striking arc of dust and gas connecting sources~A and B, make IRAS\,16293 a very complex system, where the observed gas line emission can only be explained by a combination of multiple physical components. We aim to disentangle these structures in our new observations and to map the physical origins of the observed molecular emission lines. 

This paper is based on observations from the ALMA band 7 segment of the Protostellar Interferometric Line Survey (PILS\footnote{\url{http://youngstars.nbi.dk/PILS}}) targeting IRAS\,16293. The many thousands of line detections in the PILS data set facilitate the discovery of complex organic molecules ($>$6 atoms, including carbon; \citealt{herbst_vandishoeck2009}) and the study of their interstellar chemistry. Relatively weak lines of complex molecules are most easily separable in spectra extracted toward spatial positions near the kinematically simple and spatially compact source~B, with narrow ($\sim$1~\kms), single-moded spectral line shapes. Most analysis based on the PILS data cubes has therefore been restricted to spatial positions close to the two protostars \citep[e.g.,][]{jorgensen2016,coutens2016,lykke2017,ligterink2017,fayolle2017,coutens2018b}. In contrast, in this work, we fully exploit the spatial dimensions of the PILS data set, but focus on a few restricted frequency ranges containing lines of well-known, simple molecular species. The aim is to study the kinematic structure of the binary system and the intervening and surrounding gaseous material. 

We summarize the characteristics of the observational data in Sect.~\ref{sec:obs}, and describe the selection of molecular tracers and the morphology and dynamics observed in each of them in Sect.~\ref{sec:results}. Section~\ref{sec:analysis} presents analysis of the physical characteristics of the bridge filament between the two protostars and the kinematics of outflow motions. Discussion of the results is given in Sect.~\ref{sec:discussion}, and Sect.~\ref{sec:conclusions} summarizes the main conclusions.

\begin{figure*}
	\resizebox{\hsize}{!}{\includegraphics{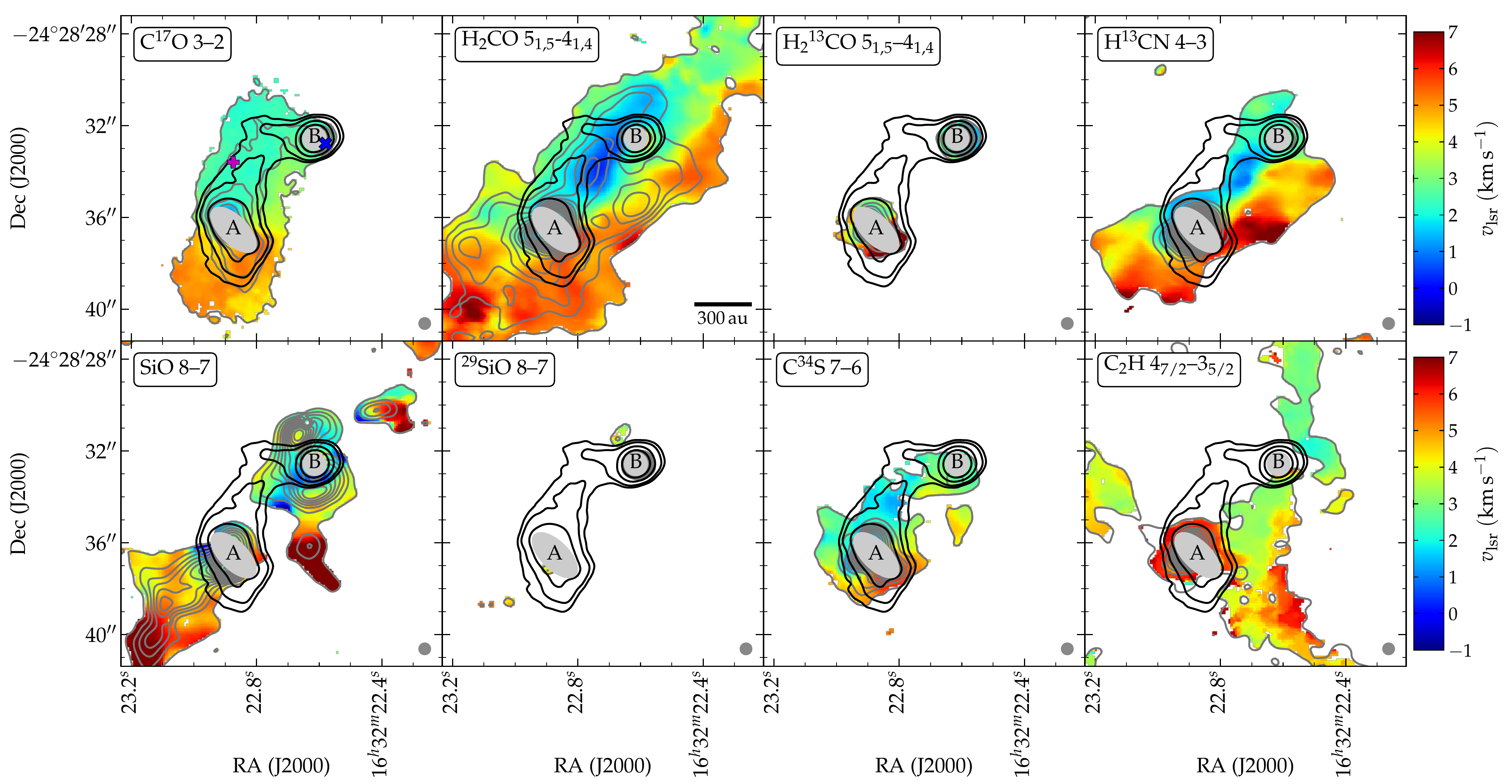}}
	\caption{Velocity maps of molecular species listed in Table~\ref{t:selectedlines} (except $^{12}$C$^{16}$O, see Fig.~\ref{fig:highvelocityCO}). Black contours represent 0.87~mm dust continuum at levels of 30, 45, 100, 250 m\Jyperbeam. 
	Gray contours for integrated spectral line intensity start at 0.35 \Jyperbeam\,\kms, and velocity values are only shown where the integrated intensity is above this threshold. The threshold for \CCH\ is 0.07 \Jyperbeam\,\kms, to reflect its narrower integration range ([$-1,+7$] instead of [$-4$,$+11$]~\kms); it is 0.20 \Jyperbeam\,\kms\ for $^{29}$SiO. Higher level integrated line intensity contours (also in gray) cover 14 linear steps up to the maximum intensity in each map (in units of \Jyperbeam\,\kms): 5.62 for \CstO, 21.67 for \HHCO, 8.10 for \HHthCO, 32.55 for \HthCN, 13.23 for SiO, 2.91 for $^{29}$SiO, 6.54 for \CtfS, and 6.02 for \CCH. For comparison with the contour levels: typical rms noise levels in emission-free regions of the integrated intensity maps are 0.03--0.06 \Jyperbeam\,\kms.
	Velocities between +7 and +11~\kms\ are represented by the darkest red color, and velocities between $-4$ and $-1$~\kms\ by the darkest blue color.  
	The circular synthesized beam of 0.5\arcsec\ in FWHM is indicated in the bottom right of all panels, except the \HHCO\ panel, which features a scale bar representative of 300~au at the source distance of 120~pc. 
	The blue `$\times$' sign in the top left panel marks the position 0.5\arcsec\ southwest of the continuum peak of source~B, often used to extract signals of complex organic molecules \citep[e.g.,][]{coutens2016,lykke2017,ligterink2017}. The magenta `$+$' sign marks the position at which the representative bridge filament spectral profiles are extracted (see Appendix~\ref{sec:contamination}).  
	 }
	\label{fig:velocitymaps}
\end{figure*}

\begin{figure*}
	\resizebox{\hsize}{!}{\includegraphics{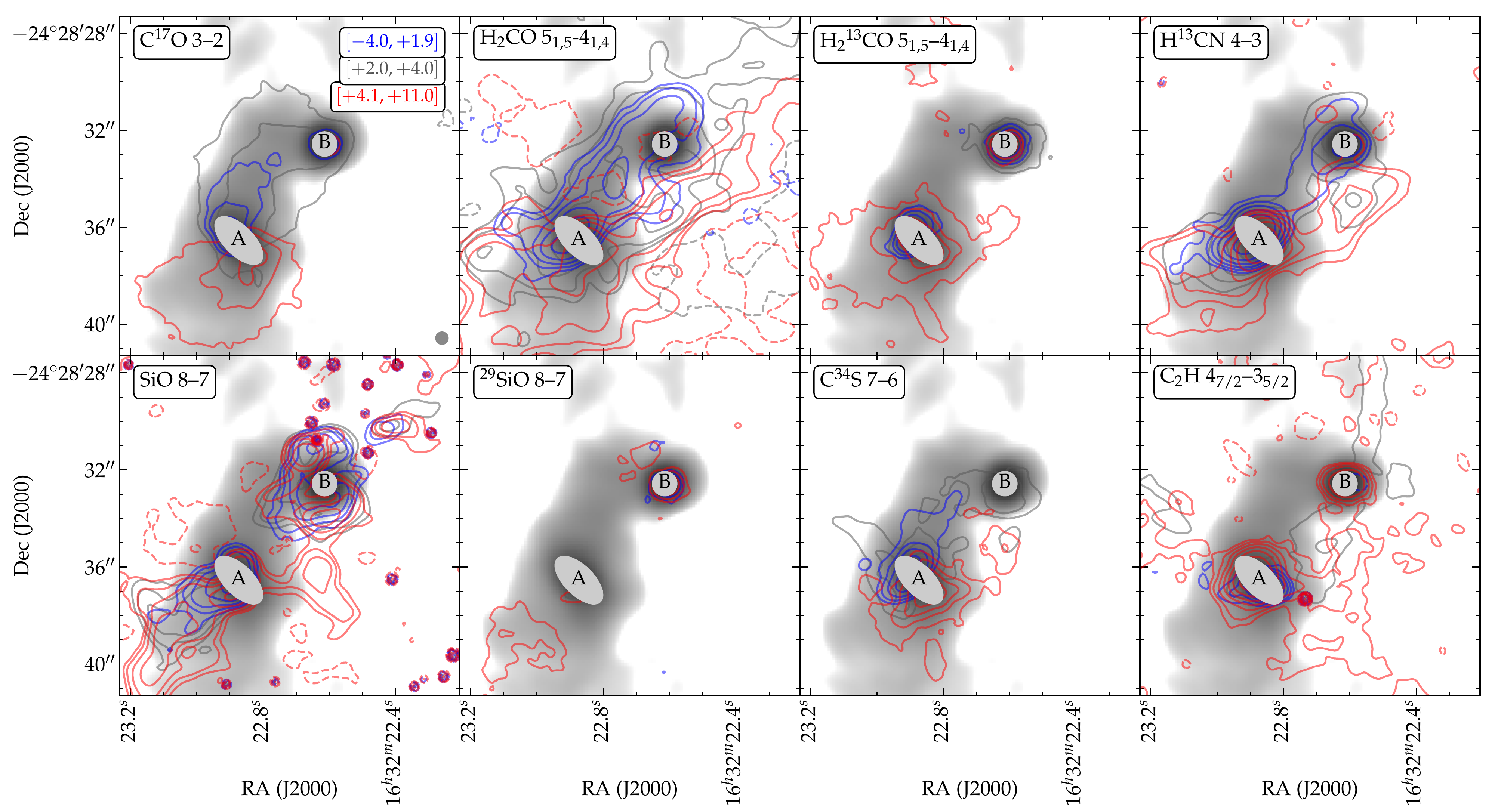}}
	\caption{Integrated velocity ranges (blue, gray, and red contours) for each molecular line tracer. Contour levels start from 0.2 \Jyperbeam\,\kms\ (equivalent to 3.3--7$\sigma$), except for \HHCO\ (0.4 \Jyperbeam\,\kms, rms noise levels poorly quantified due to lack of emission-free regions), \HHthCO\ (0.1 \Jyperbeam\,\kms, $\sim$3.5--5$\sigma$), $^{29}$SiO and \CCH\ (both 0.07 \Jyperbeam\,\kms, $\sim$1.6--3.5$\sigma$), and increase by a factor of two to the next level. Equivalent negative contours are plotted in dashed style. 
	The 0.87~mm continuum emission is shown in grayscale, stretching from 0.002 to 2.0 \Jyperbeam.
		We note that in the SiO 8--7 panel, the scattered, compact, negative value contours with angular sizes less than one beam are not real signal, but artefacts from the continuum subtraction process.
	} 
	\label{fig:all_blue-syst-red}
\end{figure*}

\begin{figure}
	\resizebox{\hsize}{!}{\includegraphics{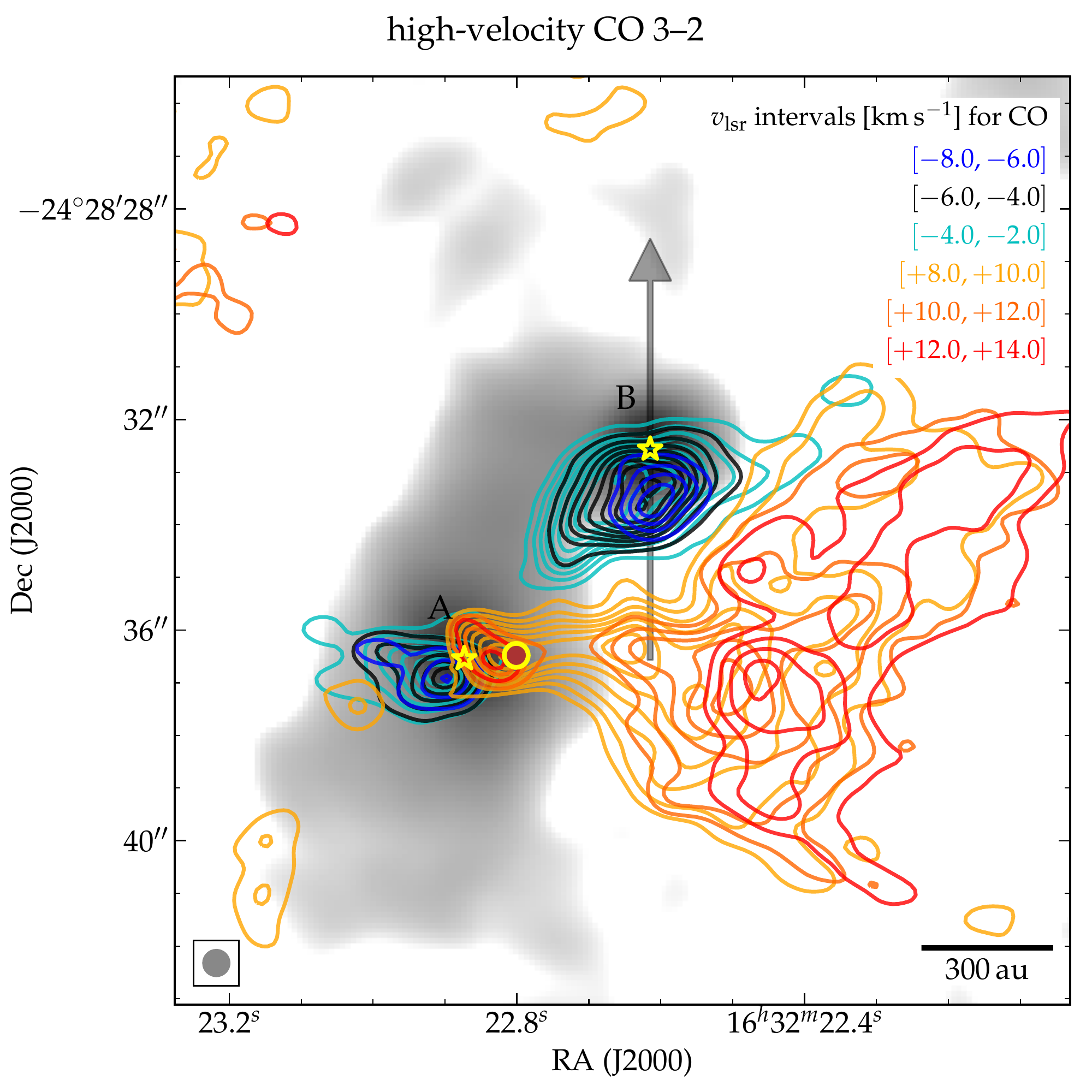}}
	\caption{
	Map of six different velocity bins of the main CO isotopologue, highlighting high-velocity gas at $>5$~\kms\ from the systemic velocity (\vlsr$\approx$$+3$~\kms) of the protostars: absolute velocity difference of [5,7]~\kms\ in cyan and orange contours, of [7,9]~\kms\ in black and darker orange, and of [9,11]~\kms\ in blue and red. The lowest contour level and step to the next contour is 1.0 \Jyperbeam\,\kms. The 0.87~mm dust continuum is shown in grayscale. The arrow indicates the slice for the position-velocity diagram shown in Fig.~\ref{fig:pv_sourceB}. The position of the water maser spot at \vlsr=$+6.1$\,\kms\ \citep{dzib2018} is marked with a red circle with a yellow outline, $\sim$1\arcsec\ west of source~A. 
	A linear scale indicator is plotted in the bottom right corner, and the size of the synthesized PILS beam in the bottom left corner. 
	}
	\label{fig:highvelocityCO}
\end{figure}

\begin{figure}
	\resizebox{\hsize}{!}{\includegraphics{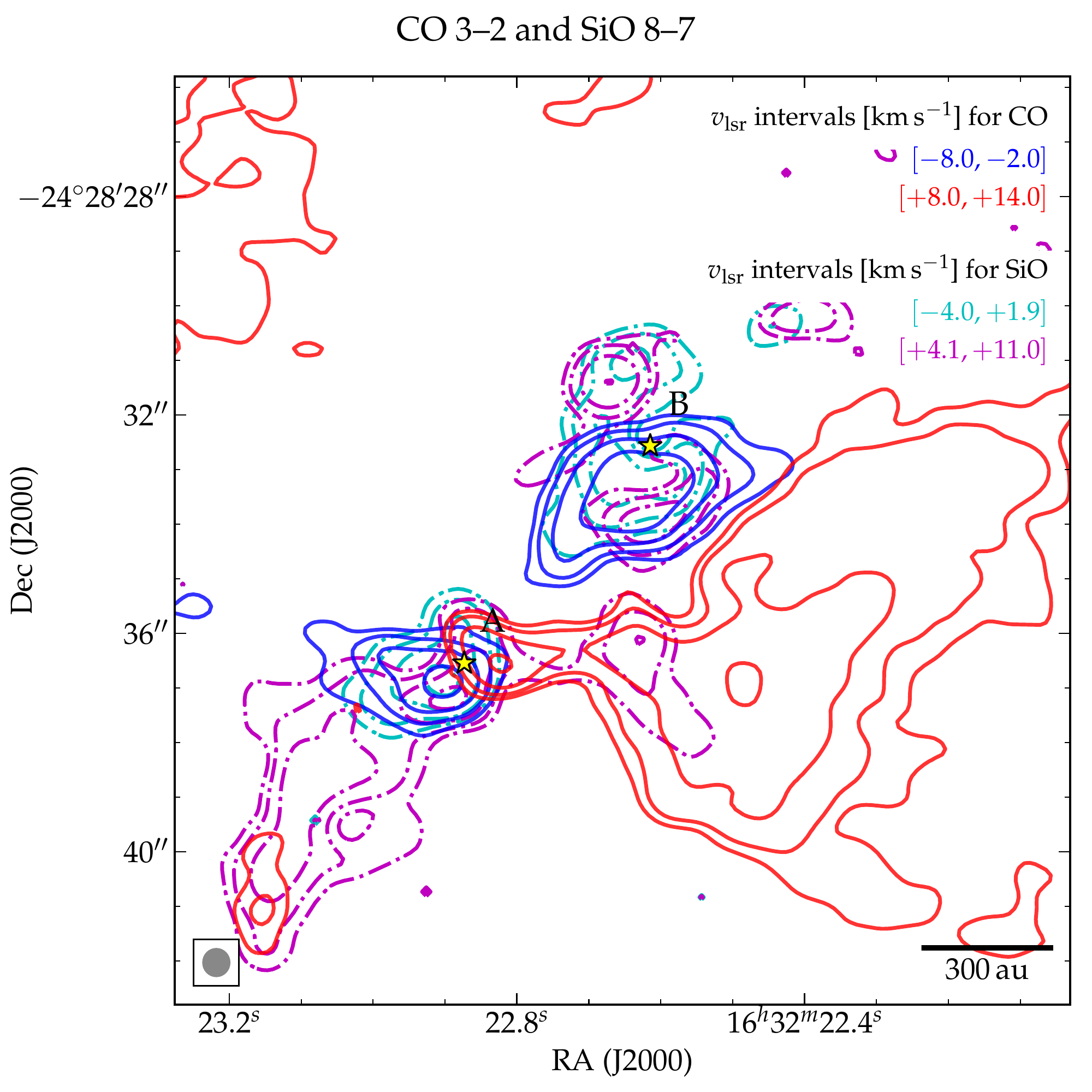}}
	\caption{Overlay of the maps of high-velocity CO emission (solid blue and red contours), and non-systemic SiO emission (dash-dotted cyan and magenta contours). The three separate CO velocity bins on either side of the systemic velocity shown in Fig.~\ref{fig:highvelocityCO} are collapsed into one bin for this figure (solid blue and red contours). The blueshifted and redshifted SiO emission (dashed cyan and magenta contours) is the same as shown in the bottom left panel of Fig.~\ref{fig:all_blue-syst-red}. 
	A linear scale indicator and the size of the synthesized PILS beam are plotted in the bottom right and bottom left corners, respectively.  }
	\label{fig:COandSiO}
\end{figure}

\section{Observations}
\label{sec:obs} 

The PILS program was conducted with ALMA between June 2014 and May 2015, using both the main array of 12-m dishes and the shorter baselines available in the array of 7-m dishes in the Atacama Compact Array (ACA). Its end product comprises a three-dimensional data set covering the uninterrupted spectral range 329--363~GHz (all in ALMA band 7) in spectral channels of 0.244~MHz in width. The sky area of $\sim$15\arcsec\ in diameter is covered in a single pointing with the primary beam of the \mbox{12-m} dishes, sampling the two binary components of IRAS~16293 and their surroundings with a synthesized beam FWHM of 0.5\arcsec. Self-calibration of the visibility phases was applied, based on the continuum signal. While the main array provides the unprecedented sensitivity, which is roughly uniform at $\sim$8 m\Jyperbeam\,channel$^{-1}$ across the full spectral range, the addition of ACA baselines ensures that large scale emission up to $\sim$13\arcsec\ is recovered. 
We refer to \citet{jorgensen2016} for additional details of observing conditions, data characteristics, and the data processing strategy.

\section{Results: observed dynamics and morphology}
\label{sec:results}

The morphology of the dust continuum emission in IRAS\,16293 as observed with ALMA at 0.5\arcsec\ resolution at 0.87~mm (\citealt{jorgensen2016}; \citetalias{jacobsen2018}) can be broken up into three distinct components: a nearly circular structure that hints at a face-on disk surrounding source~B; an elliptically shaped structure related to the inclined disk-like structure around source~A; and a ridge of material stretching from B to A and even beyond, to the southeast of A. The centroid coordinates of the continuum peaks of components A and B, measured using two-dimensional Gaussian fits on the PILS 0.826--0.912~mm continuum image, are listed in Table~\ref{t:peakcoordinates}. These positions of the 0.87~mm continuum measured from our PILS data products are consistent with the 1.15--1.30~mm continuum peak locations reported by \citet{oya2018} to within 0.05\arcsec, that is, one tenth of the synthesized beam size of either observation. 

\begin{table}
\centering 
\caption{Coordinates of submillimeter continuum peaks in IRAS\,16293$-$2422.}
\label{t:peakcoordinates} 
\begin{tabular}{l l l l l l }
\hline\hline
Component			& Right Ascension 		& Declination \\
					& (J2000)				& (J2000) \\
\hline
source~A	& 16\h32\m22\fs873 		& $-$24\degr28\arcmin36\farcs54 \\
source~B	& 16\h32\m22\fs6147	& $-$24\degr28\arcmin32\farcs566 \\
\hline 
\end{tabular} 
\tablefoot{The uncertainties on fitted positions are 0.03\arcsec\ for both coordinates of component A, and 0.005\arcsec\ for component B.}
\end{table}

\begin{table*}
\centering 
\caption{Selected line transitions.}
\label{t:selectedlines} 
\begin{tabular}{l l l l l l }
\hline\hline
Molecule	& Transition			& Rest frequency\tablefootmark{a} & \Eup/$k$ & $n_\mathrm{crit}$\tablefootmark{b}	& Traced component\tablefootmark{f}	\\ 
		&					& (GHz)		& (K)		& (\pccm)			& \\
\hline
Dust		& (continuum)			& 344 (0.87~mm) & -	& 	-	& Quiescent bridge between A and B. \\
\CstO	& $J$~=~3--2				& 337.061\tablefootmark{c}		&  32.5	& \pow{4.1}{4}	& Quiescent bridge between A and B. \\
$^{12}$C$^{16}$O $\tablefootmark{d}$ & $J$~=~3--2		& 345.796		& 33.2	& \pow{4.1}{4}	& Redshifted, outflow cone accelerating westward from A; \\ 
		& 					& 			& 		& 								& onset of eastern counterpart of westward outflow; \\ 
		& 					& 			& 		& 								& blueshifted knot south of B. \\
\HthCN	& $J$~=~4--3			& 345.33977 	& 41.4					& \pow{1.2}{8}	& Redshifted northern wall of PA 270\degr\ outflow; \\
		&					&			&		&								& Redshifted outflow southeast (PA 135\degr) from A. \\
o-\HHCO	& $J_{K_a,K_c}$=~5$_{1,5}$ -- 4$_{1,4}$	& 351.76864	& 62.5	& \pow{7.5}{7}			& Redshifted northern wall of PA 270\degr\ outflow; \\ 
		& 					&			&		&								& Blueshifted interface wall with bridge. \\
		&					&			&		&								& Redshifted outflow southeast (PA 135\degr) from A. \\
o-\HHthCO & $J_{K_a,K_c}$=~5$_{1,5}$ -- 4$_{1,4}$	& 343.32571	& 61.3	& \pow{7.5}{7}		& Redshifted northern wall of PA 270\degr\ outflow; \\ 
\CtfS		& $J$~=~7--6				& 337.39646	& 50.2	& \pow{1.5}{8}					& Blueshifted gas in same direction as `axis' of bridge. \\
SiO 		& $J$~=~8--7				& 347.33058	& 75.0	& \pow{6.1}{7}					& Axes of of redshifted (PA 135\degr) and \\
		&					&			&			&							& \ \ blueshifted (315\degr) outflows from A; \\
		&					&			&		&								& Mixed-velocity pockets north and south of B; \\
$^{29}$SiO	& $J$~=~8--7			& 342.98084	& 74.1	& \pow{5.9}{7}	& Redshifted outflow southeast (PA 135\degr) from A.  \\ 
		&					&			&		&								& Mixed-velocity pockets north and south of B; \\
C$_2$H		& $N_J$~=~$4_{7/2}$ -- $3_{5/2}$ & 349.39997\tablefootmark{e}	& 41.9 & \pow{1.4}{7} & Quiescent, narrow filament across B. \\ 
\hline
\end{tabular}
\tablefoot{ 
\tablefoottext{a}{Rest frequencies taken from the Cologne Database for Molecular Spectroscopy, CDMS \citep{muller2005,endres2016}. Original sources of spectroscopic data: \CstO, \citet{klapper2003}; CO, \citet{winnewisser1997}; \HthCN, \citet{cazzoli2005}; \HHCO\ and \HHthCO, \citet{cornet1980}; \CtfS\ \citet{gottlieb2003}; SiO and $^{29}$SiO, \citet{muller2013}; \CCH, \citet{padovani2009}. } \\
\tablefoottext{b}{Critical density is calculated as the ratio of the Einstein $A$ coefficient and the collision rate at a temperature of 100~K, which are accessed through the LAMDA database \citep{schoeier2005}, with the following original sources: CO (adopting o-\HH\ as dominant collision partner), \citet{yang2010}; \HthCN, \citet{dumouchel2010} for main isotopologue; \HHCO\ and adopting identical rates for \HHthCO, \citet{wiesenfeld2013}; \CtfS\ adopting rates for the main isotopologue C$^{32}$S, \citet{lique2006}; SiO and adopting identical rates for $^{29}$SiO, \citet{dayou2006}; \CCH, \citet{spielfiedel2012}.} \\
\tablefoottext{c}{The hyperfine splitting of \CstO~3--2 and \HthCN~4--3 is discussed in Appendix~\ref{sec:contamination}.} \\
\tablefoottext{d}{For $^{12}$C$^{16}$O, we study morphological features using only velocity channels at $|$ \vlsr $-$ $v_\mathrm{systemic}$ $| > 5$~\kms, as shown in Fig.~\ref{fig:highvelocityCO}.} \\
\tablefoottext{e}{The rest frequency for \CCH~$4_{7/2}$ -- $3_{5/2}$ is the average of the tabulated rest frequencies for the individual hyperfine components, $F$=4--3 at  349.39928 GHz and $F$=3--2 at 349.40067 GHz.}\\
\tablefoottext{f}{source A and source B are abbreviated as `A' and `B'; `PA' denotes position angle east of north, following Fig.~\ref{fig:outflowcartoon}.}
} 

\end{table*}

We select spectral signatures of \CstO, \HthCN, \HHCO, \HHthCO, \CtfS, SiO, $^{29}$SiO, and \CCH. 
The motivation for selecting these tracer transitions is that their emission is spatially extended across the inter- and circumbinary region, the molecules are chemically simple, and are chosen because they probe a range of densities, temperatures (see \Eup\ and $n_\mathrm{crit}$ in Table~\ref{t:selectedlines}) and physical processes such as shocks. \CeiO\ 3--2 largely follows the spatial distribution of \CstO\ \citepalias{jacobsen2018}, and the former is not further discussed in this paper. Although we recognize the additional value of studying other tracer molecules also covered in the 329--363~GHz PILS data, the selection in this paper is deliberately restricted to its current scope. 
The rest frequencies for the relevant transitions of these species are listed in Table~\ref{t:selectedlines}. For the systemic velocities of each of the protostars, we adopt the statistically averaged values based on the multiline analysis by \citet{jorgensen2011}. 
Velocity channel maps of each of these species are shown in Figs.~\ref{fig:C17Ochannelmaps}--\ref{fig:CCHchannelmaps}, with a velocity range in \vlsr\ between 0.0 and $+6.5$~\kms, bracketing the systemic velocities of $+3.2$ and $+2.7$~\kms\ for sources A and B in the IRAS\,16293 binary \citep{jorgensen2011}. As our aim is to study material nearby, but not in the dense protostellar sources, two masks are placed on the disks of each source in all maps presented in this work. This procedure ensures that high-intensity, partially (self-)absorbed spectral line signals do not skew the intensity scaling on the integrated intensity maps and the calculation of the weighted velocity (moment 1) maps. 
Contamination of the selected molecular line transitions by nearby transition of other species are discussed in detail in Appendix~\ref{sec:contamination}. 
Spectral line shapes deviate from Gaussian profiles in some cases, particularly in directions where line optical depth is high, and/or outflowing or infalling gas motions contribute to the gas emission.

We construct velocity maps of the selected species by considering only channels with \vlsr\ in the range [$-4$,$+11$]~\kms\ (except for \CCH, where we choose [$-1$,$+7$]~\kms\ to limit contamination by other species), and calculating an intensity-weighted mean velocity for each pixel, excluding all flux density values below 40 m\Jyperbeam\ (i.e., roughly 4--6 $\sigma$ depending on the noise level in the particular section of frequency coverage). The result is shown in Fig.~\ref{fig:velocitymaps}. The velocity structure of each tracer is also displayed in the form of contour maps for three integrated ranges of velocity (red, systemic, blue) in Fig.~\ref{fig:all_blue-syst-red}, and split out for each 0.2~\kms\ spectral channel in Appendix~\ref{sec:channelmapfigures}. 

As seen in Fig.~\ref{fig:velocitymaps}, the \CstO\ gas has an intensity-weighted mean velocity restricted to $\pm$ 1~\kms\ of the systemic velocity in the bridge region between source~A and B. The bridge is therefore kinematically quiescent, that is, we find no evidence that the bridge is part of an outflow motion nor that it is a rotating structure. The spatial distribution of \CstO\ coincides with the dust bridge traced by submillimeter continuum, while none of the other (higher density) tracers in this study have intensity peaks that are cospatial with the bridge, at 0.5\arcsec\ resolution (60~au in projection, at the distance of the source). The bridge morphology is also recovered in the PILS \CeiO~3--2 map \citepalias{jacobsen2018}, while the \thCO~3--2 starts looking markedly different due to high line optical depth \citepalias{jacobsen2018}. The \CtfS~7--6 distribution, discussed in more detailed in Sect.~\ref{sec:kinematics}, may seem to overlap partly with the dust and \CstO, but we do not regard it as tracing the bridge. 

A tentatively axisymmetric structure is seen in the channel maps of \HHCO\ and \HthCN\ (Figs.~\ref{fig:velocitymaps}, \ref{fig:all_blue-syst-red}, \ref{fig:H13CNchannelmaps}, \ref{fig:H2COchannelmaps}): two arcs emanating from source~A, one on the NE side at velocities between $+2$ and $+3.5$~\kms, and its counterpart on the SW side at velocities between $+4$ and $+5.5$~\kms\ (e.g., Fig.~\ref{fig:H13CNchannelmaps}). These arcs bracket the axis defined by the NW outflow seen in CO~6--5 by \citet{kristensen2013a}, and are kinematically symmetric around \vlsr\ = $+3$~\kms. Examining the molecular tracers considered in this work: this velocity gradient (roughly perpendicular to the main axis of the bridge) is spatially more compact in \HthCN\ than in \HHCO; there may be evidence of it in \CtfS; but it is absent in \CstO, SiO, and \CCH\ (Fig.~\ref{fig:velocitymaps}, Fig.~\ref{fig:all_blue-syst-red}). We therefore conclude that the striking symmetry observed in \HHCO\ is not the result of a bulk rotation of the gas about the axis of the bridge or the axis of the northwest outflow from source~A. 

While none of the tracer molecules listed above have signals bright enough to study faint line wings at velocity offsets more than 4~\kms, the main isotopologue of CO provides sufficiently bright signal to probe higher line-of-sight velocities. In fact, in Fig.~\ref{fig:highvelocityCO} we use only high-velocity ($>$5~\kms) channels of CO 3--2 emission, thereby avoiding complications in interpreting extremely optically thick emission at lower velocities. The western lobe of the east-west outflow pair is prominently visible in CO 3--2 in a cone-shaped distribution, with a much smaller region of blueshifted counterpart appearing on the eastern side of source~A. At somewhat coarser angular resolution (1.5\arcsec\ beam, using the SMA), \citet{yeh2008} have previously studied the east-west outflow from source~A in CO~2--1 and 3--2. Compared with their work, our 0.5\arcsec\ resolution map in CO~3--2 reveals a more cone-like shape of the westward, redshifted outflow lobe, and there is an apparent acceleration taking place with increased distance from source~A. See Sect.~\ref{sec:kinematics} for a discussion of the structures described here.   
There is also bright CO 3--2 emission with an emission peak 1\arcsec\ south of source~B, with velocities ranging up to 10~\kms\ blueshifted with respect to source~B. Its location is consistent with the position `b2' described in \citet{yeh2008}. In our map (Fig.~\ref{fig:highvelocityCO}), this blueshifted knot is clearly morphologically connected to source~B. 

We observe that the SiO velocity map in Fig.~\ref{fig:velocitymaps} traces both sides of the NW-SE outflow pair extending up to $\sim$8\arcsec\ from source~A, blueshifted on the northwest side and redshifted on the southeast side. The lower optical depth line of $^{29}$SiO only shows emission on the redshifted side southeast of source~A. In contrast, a redshifted southeast outflow appears absent in the high-temperature tracer CO 6--5 \citep{kristensen2013a}, although it is seen in colder gas (CO 2--1; \citealt{girart2014}). Likewise, the PILS CO 3--2 data (Fig.~\ref{fig:highvelocityCO}) do show redshifted emission about 6\arcsec\ southeast of source~A, and it is very prominent in the SiO velocity map (Fig.~\ref{fig:velocitymaps}) at \vlsr $\geq$ $+5$~\kms, as well as in \HHCO\ and \HthCN. 
In addition, a significantly redshifted patch of emission is observed in SiO about 3\farcs5 due west of A (and 4\arcsec\ south of B), overlapping with the `kink' location in the CO outflow, as highlighted in Fig.~\ref{fig:COandSiO} (see also Sect.~\ref{sec:kinematics}). SiO 8--7 is also seen prominently 1--2\arcsec\ north and south of source~B, with mixed red and blue velocity components along the line of sight (Figs.~\ref{fig:all_blue-syst-red}, \ref{fig:SiOchannelmaps}), as well as in the fainter $^{29}$SiO line (Fig.~\ref{fig:29SiOchannelmaps}). This structure is consistent with that observed in SiO~6--5 by \citet{oya2018}. 

Finally, a filamentary structure connected to source~B with a position angle of $\sim$170\degr\ is visible in \CCH, but not in any other tracer. Its velocity structure is flat, restricted to \vlsr\ values between $+2$ and $+4$~\kms\ (see Sect.~\ref{sec:kinematics} for further discussion). The redshifted (>$+5$~\kms) emission in the velocity map of \CCH\ (Fig.~\ref{fig:velocitymaps}) is due to a bright CH$_3$CN transition at 349.393~GHz. In positions near source~A, the CH$_3$CN line also contaminates channels closer to the systemic velocity of \CCH\ (see Appendix~\ref{sec:contamination}). In other positions in the map with integrated intensity of \CCH\ above 0.07~\Jyperbeam\,\kms, the spectral profile shows a characteristic structure with two roughly equal intensity peaks spaced by 1.4~MHz, the separation between the $F$=4--3 and $F$=3--2 hyperfine components (see Table~\ref{t:selectedlines}, Fig.~\ref{fig:CCHspectrum} in this work, and Fig.~4 of \citealt{murillo2018}). 
The interpretation of the observed morphology and velocity structure is addressed in Sect.~\ref{sec:kinematics} and Sect.~\ref{sec:discussion}.

\section{Analysis}
\label{sec:analysis}

\subsection{Radiative transfer modeling of the interbinary bridge} 
\label{sec:bridgemodel}

The bridge between protostars A and B, as traced by the submillimeter dust continuum and cold \CstO\ gas (Sect.~\ref{sec:results}), is kinematically quiescent (Sect.~\ref{sec:results}) and its central velocity at \vlsr=+3~\kms\ is consistent with the velocities of the individual protostars ($+2.7$ and $+3.1$~\kms). It is therefore unlikely to be associated with any of the outflows, and it is treated as a separate entity in this work. Besides \CstO, none of the other selected molecules match, simultaneously, the morphology of the dust bridge filament and the narrow line-of-sight velocity distribution of \CstO. 

In principle, an assessment of density and temperature conditions throughout the inter- and circumbinary region of IRAS\,16293 could be set up using spectral line intensity ratios of various molecules and transitions highlighted in this work (see Table~\ref{t:selectedlines}). However, given the multitude of partially resolved dynamical components, it is not at all certain that line intensities measured toward a particular position emanate in the same gas for one transition of a species and another transition of another species. We therefore refrain from embarking on such an analysis. 

%
%
To study the quiescent interbinary bridge, we use a curved cylinder filament model, spanning 636\,au between and enveloping the two sources in an arc-like structure. The density profile of the bridge arc is a power-law which decreases with radius to the power $-0.25$ and is independent of the distance from either protostellar source. The enclosing envelope extends to beyond our field of view, and has a purely spherical geometry, with a density profile decreasing radially with a power of $-1.7$ and a density plateau imposed at radii within 600 au. This is the same filament model used in \citetalias{jacobsen2018}, who show that, when using luminosities of 18 $L_{\odot}$ and 3 $L_{\odot}$ for sources~A and B, respectively, the model yields a qualitative match with the morphology and peak emission levels seen in the 0.87~mm dust continuum emission and \CeiO/\CstO\ gas line emission. The fiducial model adopted in this paper is equivalent to `Rotating Toroid model 1' in \citetalias{jacobsen2018}. The temperature structure of each cell in the model is derived self-consistently, based on the adopted luminosities of both sources~A and B (Fig.~F.4 of \citetalias{jacobsen2018}). 
We do not scale distances in the model to the newly published distance of IRAS\,16293 (see Sect.~\ref{sec:intro}), to allow inclusion of and direct comparison with results from the modeling work of \citetalias{jacobsen2018}. An 18\% increase in source distance would have little impact on the model outcomes, compared with other assumptions with considerably larger uncertainties, such as dust opacity and sublimation thresholds (see below). 

The line radiative transfer code \texttt{LIME} {\citep{brinch2010} is used to obtain synthetic line emission cubes of the \HHCO\ 5$_{1,5}$--4$_{1,4}$ and \HthCN~4--3 gas line emission, analogous to the strategy used in \citetalias{jacobsen2018} for CO and its isotopologues. We use an abundance jump model to emulate freeze-out onto the dust grains. When the dust temperature is below the sublimation temperature
, the gas phase abundance is decreased by a depletion factor. See Table \ref{t:lime} for more information on the individual molecules. Uncertainties in the sublimation thresholds could be up to $\sim$30\% (see references in Table~\ref{t:lime}). This study only considers the inter- and circumbinary region and excludes the warm `disk regimes' at $\gtrsim$50\,K (cf.~the model temperature structure in Fig.~F.4 of \citetalias{jacobsen2018}). Therefore, adjusting the adopted sublimation temperature by 30\% would only impact the modeled emission morphology of CO species, and only those that are optically thin (\CstO, \CeiO), while those of \HHCO\ and \HthCN\ would be left unaffected. The number density of \HH\ is above $10^8\,$\pccm\ in the modeled bridge filament, sufficiently high to validate the occurrence of depletion of gas-phase molecules onto dust grain surfaces. Since we are investigating the appropriateness of a static filament structure to the observation of the quiescent bridge only, we keep the velocity structure of the model static, with only random velocity dispersions included, such as from turbulence. We fix the built-in velocity dispersion parameter (Doppler $b$) in \texttt{LIME} to 1~\kms. This value is the 1/e half-width, equivalent to a FWHM of 1.4\,\kms\ for a Gaussian line profile, which is representative of the measured line widths in the region under study. The only free parameter is the molecular abundance. Twenty different abundance values are run for each molecule in the range given in Table~\ref{t:lime}. After \texttt{LIME} produces a synthetic line emission cube, it is convolved with a $0.5^{\prime\prime}\times0.5^{\prime\prime}$ beam in the image analysis software package \texttt{MIRIAD}. 
We refer to \citetalias{jacobsen2018} for further details of the model definition and the radiative transfer approach. 


\begin{table}
  \caption{\texttt{LIME} model parameters. } 
  \centering
 \begin{tabular}{@{}l@{\ }c@{\ }c@{\ }c@{\ }c@{}}
  \hline\hline
  Molecule & $X$/H$_2$ & $T_{\mathrm{subl}}$\tablefootmark{a}  & Depletion & $b$\tablefootmark{c} \\ 
  		& 	range	& 		[K]		& factor\tablefootmark{b} & [\kms]  \\
  \hline 
  \CstO	& \pow{5.58}{-8} & 30	& 100 & 1 \\
  o-\HHCO & [$1\times10^{-8}$, $3\times10^{-6}$] & 50 & 100 & 1\\
  H$^{13}$CN & [$2\times10^{-12}$, $2\times10^{-10}$] & 100 & 100 & 1\\
  \hline
  \end{tabular}
  \tablefoot{
  All parameters for \CstO\ are identical to those in \citetalias{jacobsen2018} (see their Table~3).\\ 
  \tablefoottext{a}{Ice mantle sublimation temperature for HCN follows those of other CN-bearing species \citep[e.g.,][]{noble2013a}. That of \HHCO\ follows \citet{aikawa1997}, \citet{ceccarelli2001} and \citet{rodgers2003}. }\\
    \tablefoottext{b}{When depletion is applied in the model, the gas phase abundance of the molecule in question is divided by the depletion factor.}\\
  \tablefoottext{c}{Doppler parameter as defined in \texttt{LIME}, that is, the 1/e half-width of a thermally broadened line.} \\
  } 
\label{t:lime}
\end{table}

%
%

\begin{figure}
	\resizebox{\hsize}{!}{\includegraphics{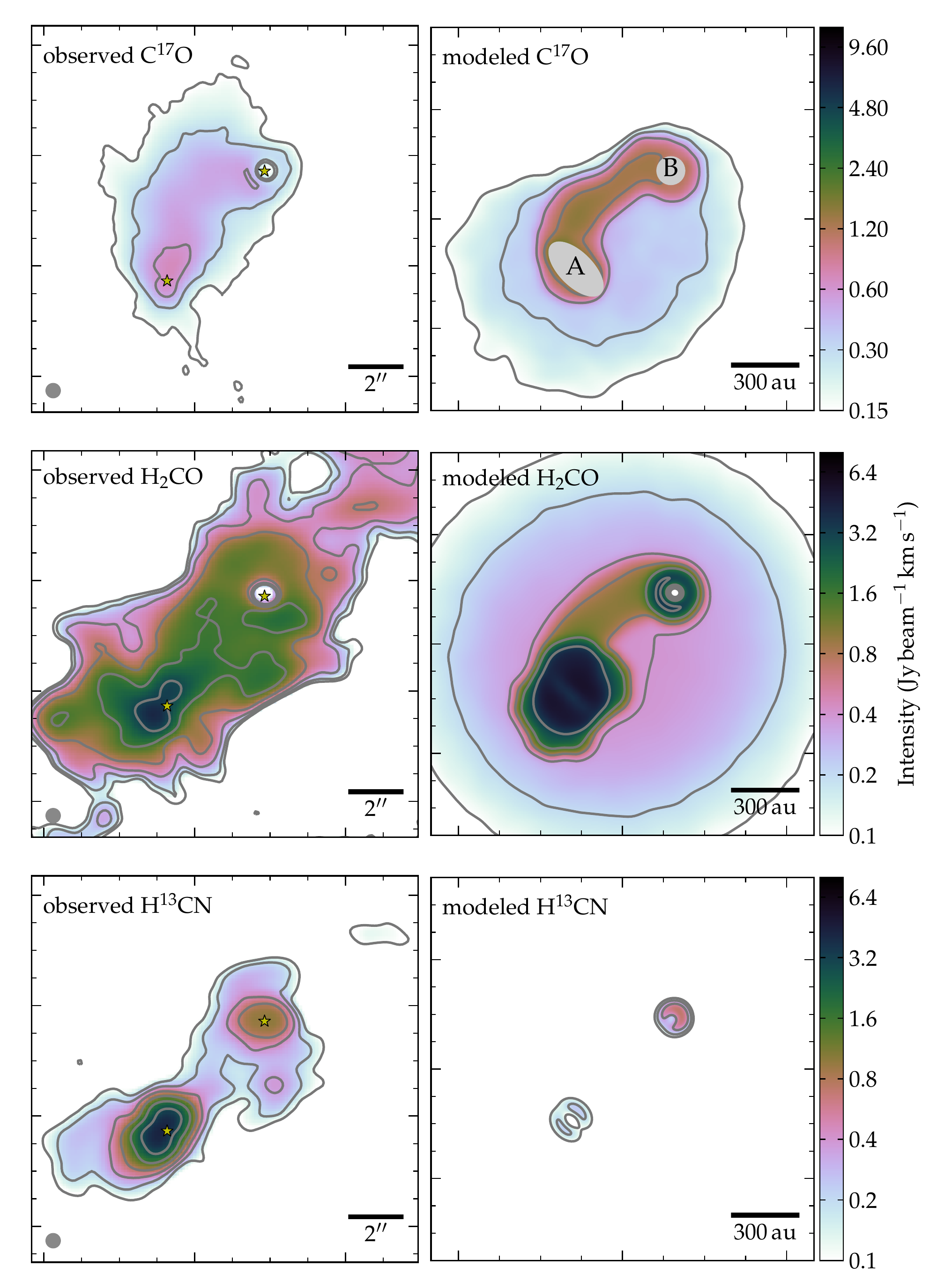}}
	\caption{Comparison of observed (left) and modeled (right) morphology of spectral line emission from \CstO, \HHCO, and \HthCN. In the \CstO\ model panel, the disk domains are masked (as in Fig.~\ref{fig:velocitymaps}). Total intensity from the observed cubes is calculated by integrating over \vlsr\ range [$+2$,$+4$]~\kms. The continuum peak locations are marked by star symbols, the ALMA interferometric beam size is indicated in the bottom left of each `observed' panel. The modeled line intensity maps are convolved with the observational beam size, to aid direct comparison. Line intensity is represented by the color scale and gray contours (identical for observed and modeled panels: lowest level at 0.15 for \CstO, 0.10 Jy/beam\,\kms\ for \HHCO\ and \HthCN, and increasing by a factor of 2 each level). Black contours indicate the 0.87~mm dust continuum.}
	\label{fig:obs-vs-model}
\end{figure}

In Fig.~\ref{fig:obs-vs-model}, integrated intensity maps of the radiative transfer calculations described above are juxtaposed with their observed counterparts. The latter are produced by integrating over the narrow velocity range \vlsr=[+2.0,+4.0]~\kms\ (i.e., all channels with contours in black in Figs.~\ref{fig:C17Ochannelmaps}--\ref{fig:H2COchannelmaps} in the Appendix), with the aim of filtering out dynamically active gas associated to the various outflow components which are not included in the density distribution of the model \citepalias{jacobsen2018}. As already concluded by \citetalias{jacobsen2018}, the model in \CstO\ provides a qualitative match to the morphology of the bridge connecting the two protostars, although the observed \CstO\ filament is laterally $\sim$3 times wider than the modeled curved filament. Absolute flux values produced by the model are a factor $\sim$2 higher than observed. \HHCO\ is also present and detected in its 5$_{1,5}$--4$_{1,4}$ transition across the bridge area outlined by the continuum contours, but its observed morphological shape is different from that of \CstO\ and the continuum. The difference is partly due to the inclusion of outflow-impacted gas, despite the attempt to exclude this by using a narrow velocity integration range. 
The modeled \HHCO\ emission extends radially over many hundreds of au surrounding the binary, an effect of \HHCO\ molecules being present in the gas phase in the spherical envelope component of the model. The superposed density structure of the disks-bridge-envelope model is dominated by that of the spherical envelope model at such large radii of $\sim$700 au (from the midpoint between A and B). Outside this radial distance of 700~au, the number density of the modeled envelope component drops below $10^7$~\pccm\ (the outermost gray, circular contour in Fig.~F.4 of \citetalias{jacobsen2018}). Along with radially decreasing temperatures, this leaves conditions insufficient to excite o-\HHCO~5$_{1,5}$, with a critical density of \pow{7.5}{7}~\pccm\ and an upper energy level of \Eup/$k$=62.5~K (Table~\ref{t:selectedlines}). 
In addition to tracing the bridge filament, the \HHCO\ model map shows enhanced \HHCO\ emission up to $\sim$2\arcsec\ above the disk of source~A (northwest and southeast of A), where the temperature is sufficiently high to sublimate \HHCO\ from dust grains (cf.~55~K contour in Fig.~F.4 of \citetalias{jacobsen2018}). Finally, the observed morphology of \HthCN\ in the inter-binary region does not follow the shape of the bridge filament. Most of the \HthCN~4--3 emission shown in Fig.~\ref{fig:obs-vs-model}, bottom left, is due to contamination from outflow-impacted components (despite our effort to exclude these using a narrow velocity integration interval), none of which are included in the physical model. As expected from its high critical density, the radiative transfer model does not populate the $J$=4 level of HCN species beyond the disk domains. The exception is a small region immediately above the disk plane, where \Tkin$>$100\,K (again, see Fig.~F.4 of \citetalias{jacobsen2018}), where the observed enhancement in \HthCN\ emission is reproduced by the model. 

In conclusion, the bridge morphology is observed in dust, \CstO, and partly also in o-\HHCO, and the physical model constructed by \citetalias{jacobsen2018} roughly reproduces the morphology of the dust continuum and \CstO\ emission. The partial overlap between the observed \HHCO\ emission and that produced by the model (Fig.~\ref{fig:obs-vs-model}, middle panels) should not be over-interpreted. As shown in Figs.~\ref{fig:velocitymaps} and \ref{fig:H2COchannelmaps}, the \HHCO\ observed in the direction of the bridge filament domain is largely at line-of-sight velocities that are offset from the quiescent \CstO\ bridge filament at the systemic velocity. In fact, the lower optical depth tracer \HHthCO\ in the same transition (Fig.~\ref{fig:H2[13]COchannelmaps}) shows no detectable emission in the bridge domain in the \vlsr\ range from +2.0 to +4.0~\kms. This means that the observed \HHCO\ emission is not tracing the bulk gas in the quiescent bridge filament, but rather surface layers of dynamically stirred components. 
In our interpretation, this indicates that the number densities in the modeled bridge component are higher than in reality. A more realistic model would not produce any \HHCO\ emission from the bridge component. 
\HthCN~4--3 emission is not seen outside of the disk domains in the modeled map, and does not trace the bridge morphology in the observed map (Fig.~\ref{fig:obs-vs-model}, bottom panels). The explanation for the lack of \HHCO\ and \HthCN\ emission in the bridge filament is provided by the excitation balance governed by density and temperature. Looking at the critical densities and \Eup\ values in Table~\ref{t:selectedlines}, the upper levels of the relevant \HthCN\ and \HHCO\ transitions do not get populated sufficiently. The freeze-out abundance drop in the model (Table~\ref{t:lime}) is not the cause, which we confirm with a separate radiative transfer model run in which freeze-out is neglected completely, keeping all molecules in the gas phase. In the resulting emission maps (shown in Fig.~\ref{fig:nofreezeoutmaps} of Appendix~\ref{sec:nofreezeoutmaps}), there is still no \HthCN\ 4--3 emission outside of the disk regions. Without freeze-out, the model o-\HHCO~5$_{1,5}$--4$_{1,4}$ emission map is dominated by the contribution from the cold, extended envelope (Fig.~\ref{fig:nofreezeoutmaps}). In contrast with \HthCN, \HHCO\ still does show a small intensity enhancement in part of the bridge arc region, probably reflecting its slightly lower critical density when compared with HCN~4--3. The total line-of-sight optical depth of the superposed components, particularly for \HthCN, becomes much higher in the `no freeze-out' case, as expected when all \HthCN\ molecules remain in the gas phase even in the colder, extended envelope that surrounds the binary.

The fiducial model (RTM$_1$ from \citetalias{jacobsen2018}) is not necessarily the only and best fit to the data. The only observational constraints that were taken into account were the multiwavelength SED and the distribution of CO 3--2 isotopologues and submillimeter dust emission; the \citetalias{jacobsen2018} work did not include an assessment of molecular lines tracing a range of densities. The number density at the axis of the modeled bridge arc is \pow{7.5}{8}\,\pccm. The density structure that was adopted by \citetalias{jacobsen2018} was mainly driven by the wish to reproduce the dust emission morphology. However, dust emission strength depends on several factors which are all assumed to be fixed in the model setup: dust opacity, gas-to-dust ratio, and a dust temperature being coupled rigidly to that of the gas. For example, if a different grain size distribution is assumed, higher submillimeter grain opacities could arise, lowering the peak (column) density required to match the observed dust emission. Similar arguments hold for a lower gas-to-dust ratio and a higher dust temperature. A combination of these effects can easily bring the model peak number density down by a factor of 10--100 to well below the critical densities of the CS, \HHCO\ and HCN transitions selected in this work (Table~\ref{t:selectedlines}). 
In this work, a suite of molecular tracers with different critical densities are studied, among which \CstO~3--2 (\pow{4}{4}\,\pccm), o-\HHCO~5$_{1,5}$--4$_{1,4}$ (\pow{7.5}{7}\,\pccm) and \HthCN~4--3 (\pow{1.2}{8}\,\pccm). Of these, \CstO~3--2 is the only species that traces the bridge morphology observed in dust continuum (see above), and, additionally, the distribution of two lines of \thfSOtwo\ observed in the PILS data cube (critical densities of \pow{3.8}{7}\,\pccm\ and \pow{8.3}{7}\,\pccm) is also confined to the disk domains of the two protostellar sources \citep{drozdovskaya2018}. Put together, this indicates that, indeed, the real density of the bridge is likely between \pow{4}{4}\,\pccm\ and \pow{3}{7}\,\pccm. 

Besides the low-density tracers \CstO~3--2 and 0.87~mm dust, all emission observed from other molecular species in the vicinity of the bridge domain can be ascribed to components with a measurable velocity along the line of sight, or their impact on more quiescent regions (see below in Sect.~\ref{sec:kinematics}). 

\subsection{Fragmentation of the bridge} 
\label{sec:bridgefragmentation}

It is conceivable that both protostellar sources A and B formed from the bridge filament through gravitational instability. In this context, we follow the analytic approach by \citet{ostriker1964} and treat the bridge as an isothermal cylinder. In this idealized scenario (far more simplified than the self-consistently derived temperature structure modeled by \citetalias{jacobsen2018} and used in our Sect.~\ref{sec:bridgemodel}), the thermal pressure in the cylinder keeps it from collapsing {radially} if the mass per unit length, $M/L$, is below a certain threshold, 
	\begin{equation}
	M/L = \frac{2 k T_\mathrm{kin}}{\mu m_0 G} . 
	\end{equation}
	This quantity depends linearly on kinetic temperature \Tkin\ and is inversely proportional to the mass of the typical particle $\mu m_0$, with $m_0$ the mass of a proton. It is the equivalent of Jeans mass in cylindrical geometry. 
Using $\mu=2.0$ for \HH\ and three different temperature values of [25, 30, 40]~K, we arrive at a stability threshold of [0.14, 0.17, 0.23] \Msun\ per 600~au (the approximate projected length of the bridge). 

We compare the mass per length stability criterion derived above with an observationally derived mass of bridge material, calculated as follows. We take the average \CstO~3--2 intensity of 600 m\Jyperbeam\ (integrated over the [+2,+4]~\kms\ velocity range), and use \texttt{RADEX} \citep{vandertak2007} non-LTE radiative transfer calculations to convert to a \CstO\ column density. Such a calculation requires a kinetic temperature and a number density of collision partners as input. To represent a natal, isothermal, cylindrical filament, we choose a temperature between the minimum temperature of the modeled bridge arc (24~K) and the average temperature in the bridge domain (41~K) (see Sect.~\ref{sec:bridgemodel} and \citetalias{jacobsen2018}). For number density, \nHH, we adopt $10^5$~\pccm, which is between the critical density of \CstO~3--2 and those of HCN~3--2, \HHCO~5$_{1,5}$--4$_{1,4}$ and \SOtwo~17$_{4,14}$--17$_{3,15}$ (Table~\ref{t:selectedlines}, Sect.~\ref{sec:bridgemodel}). 
With \Tkin=[25, 30, 40]\,K, we obtain a \CstO\ column density of [6.2, 5.4, 4.8]$\times$$10^{15}$\,\psqcm, only mildly sensitive to temperature. Using a $^{16}$O/$^{17}$O ratio\footnote{The adopted  $^{16}$O/$^{17}$O ratio is consistent with that measured in \HHCO\ toward IRAS\,16293B by \citet{persson2018}: (2.0--3.0)$\times$$10^3$.} of \pow{2.005}{3} \citep{wilson1999} and CO/\HH=$10^{-4}$, this corresponds to $N$(\HH)=[1.24, 1.08, 0.96]$\times$$10^{23}$\,\psqcm. Taking a rough half-power width of the dust bridge of 200~au, the total mass encompassed by the bridge `cylinder' area of 200$\times$600~au between sources~A and B is [0.0056, 0.0049, 0.0044]~\Msun. 

On the other hand, integrating the mass of all dust and gas in the bridge component of the model described in Sect.~\ref{sec:bridgemodel} yields a tenfold higher total mass of 0.055~\Msun. Either way, the total bridge mass values are below the stability threshold at the assumed temperatures of 25,  30, and 40~K. 
The discrepancy between these two bridge mass calculations is not surprising, given the assumptions in the dust model from \citetalias{jacobsen2018}. The dust opacity model, from \citet{ossenkopf1994}, uses a micron-sized dust distribution. In the hypothetical case that grain growth has already taken place and that the true size distribution of dust grains in the bridge peaks at sizes of 100-1000\,\micron, then opacities at 345~GHz (expressed per gram of material) can be up to a factor of 10 larger, which would mean that the mass determination from our 
radiative transfer calculations is overestimated by up to an order of magnitude. However, there is no evidence to support that the size of the dust grains in the interbinary bridge of IRAS\,16293 would deviate from the roughly micron-sized populations representative of the unprocessed interstellar medium. 
Moreover, the canonical gas-to-dust mass ratio of 100 was used, and any deviation from this value would also change the total mass estimate of the modeled filament. 

In the case presented above in which the filament mass is below the \citeauthor{ostriker1964} stability criterion, the filament would be supported against radial collapse. This scenario, in fact, is the only way in which a filament may persist long enough for it to fragment along its vertical (length) direction \citep{inutsuka1992} and form additional dense cores. However, the total mass budget available in the current bridge filament (see above) seems insufficient to form even a brown dwarf. 
Again in an idealized nearly isothermal scenario, \citet{inutsuka1997} have shown that separation between fragmented cores within a radially stable filament are on the order of four times its width. This ratio is consistent with the bridge observed in IRAS\,16293, with an observed diameter of $\sim$200~au, and separation between the cores of at least 600~au, knowing that the vector connecting sources~A and B may be inclined with respect to the plane of the sky. 
While the bridge filament present little to no gas motion along the line of sight ($\lesssim$1\,\kms), observations presented in this work and in other papers in the literature do not rule out gas flow vectors in the plane of the sky along the length of the filament, possibly regulated by the highly ordered magnetic field in the bridge \citep{sadavoy2018c}.

Finally, we emphasize that, if indeed the bridge filament has been host to sources A and B, the filament must have been considerably more massive in the past. After adding even half of the current mass of sources A and B (0.5~\Msun) to the current bridge filament mass calculated above, it is lifted to a factor few above the gravitational instability threshold. We emphasize that the interpretation of stability of the filament, sketched in this section, only holds for the filament in its current state. Radial collapse in an earlier evolutionary stage, with a different mass distribution and under different temperature conditions, is not ruled out.
To conclude, the observed filament is obviously more complex in nature than a static, cylindrically symmetric shape characterized by a single temperature. A true assessment of its gravitational stability should include, for example, turbulent and magnetic support. Any additional mechanisms to provide support against gravitational collapse would only increase the stability threshold, and the conclusions drawn in this section about the dynamical stability of the filament would thus remain unchanged. 

\subsection{Kinematics} 
\label{sec:kinematics}

Besides the bridge filament (Sects.~\ref{sec:bridgemodel}, \ref{sec:bridgefragmentation}), almost all other morphological components discussed in this paper are dynamic and can be attributed to protostellar outflow activity. Distinct components are discussed in this subsection and are summarized in Fig.~\ref{fig:outflowcartoon} and Table~\ref{t:selectedlines}. 

The CO~\mbox{3--2} velocity map in Fig.~\ref{fig:highvelocityCO} clearly traces the western lobe of the known east-west outflow pair. However, its eastern counterpart and the separate southeast-northwest pair are all much less pronounced, possibly because the eastern outflow is intrinsically dimmer (see also \citealt{yeh2008}) and the southeast-northwest pair may have lower line-of-sight velocity. In fact, the southeast outflow lobe from source~A is detected in larger spatial extent at more modestly redshifted velocities (\vlsr=[+6,+8]~\kms) than those shown in Fig.~\ref{fig:highvelocityCO}. 
The line-of-sight velocity in the western lobe increases with distance from source~A, implying an acceleration with time. This is consistent with a centrally driven outflow that would naturally accelerate as the density of the ambient envelope, and therefore the pressure, decreases with radius \citep[e.g.,][]{moriarty-schieven1988b,arce2007ppv} . 
In addition, the CO outflow in Fig.~\ref{fig:highvelocityCO} exhibits a relatively collimated morphology up to $\sim$3\arcsec\ ($\sim$350~au in projection) from source~A, at which point it abruptly fans out into a cone-like structure. This position coincides with a marked enhancement in SiO~8--7 emission (Fig.~\ref{fig:COandSiO}), possibly related to a simultaneous acceleration of the outflow (leading to SiO enhancement through dust grain erosion) and decrease in ambient pressure (leading to a wider angle CO structure). 
The position of this kink may be related to the radial extent of the rotating `inner envelope' described by \citet{oya2016}. However, these authors infer a radius of the inner envelope of 200~au, whereas a density drop would be needed beyond 300~au to explain the location of the kink in the CO outflow morphology described above. Alternatively, the opening angle of the western outflow may be restricted by higher pressure inside the quiescent interbinary bridge described in Sect.~\ref{sec:bridgemodel} of this work. Its radius (as observed in dust continuum and \CstO, Fig.~\ref{fig:velocitymaps}, top left; Fig.~\ref{fig:obs-vs-model}, top left) appears to be more congruent with the 350~au radial separation from source~A. 

Closer in to the source, at about 120 au projected distance west of source~A, a H$_2$O maser spot detected by \citet{dzib2018} seems to coincide with the interface between the dense, rotating, disk-like structure surrounding source A (blue on the west, red on the east) and the emerging redshifted outflow in the western direction (see Figs.~\ref{fig:outflowcartoon}, \ref{fig:highvelocityCO}). The direction of proper motion of the maser spot (position angle $\sim$290\degr, \citealt{dzib2018}) roughly follows the flow direction of the western outflow as seen in CO~3--2 PILS data (Fig.~\ref{fig:highvelocityCO}). This overlap supports the suggestion made by \citet{dzib2018} that the redshifted maser spot may be due to a launching shock at the base of the outflow. A similar argument may hold for a blueshifted maser spot detected to the east of source~A, although the spatial association is more ambiguous than for the redshifted counterpart \citep{dzib2018}. 

In contrast to $^{12}$C$^{16}$O, the other molecular tracers selected from the PILS data set mainly highlight the collimated northwest-southeast pair. \citet{kristensen2013a} detected only the northwest side of this outflow in CO~6--5, while \citet{girart2014} detected both sides in CO~3--2 and SiO~8--7. In addition to previously detected tracers, in this work, we also detect both symmetric counterparts in o-\HHCO, \HthCN: redshifted with respect to source~A to the southeast, blueshifted to the northwest (Fig.~\ref{fig:velocitymaps}). Blueshifted emission in o-\HHCO, \HHthCO, and \HthCN\ appears to trace the interface of the northwest outflow with the dust bridge. This observed blueshifted emission at the southwestern edge (`inside bend') of the bridge could be an indication of the outflow impacting the quiescent bridge and sweeping up material in its wake. Pockets of enhanced \HHCO\ emission may be the result of molecules being sputtered from the mantles of mildly shocked grains associated to one of the outflow lobes. Model calculations show that such effects can be attained by shock velocities of $\sim$10--15\,\kms\ \citep{caselli1997,suutarinen2014}. More energetic shocks would produce increased amounts of SiO in the gas phase through erosion of the grain cores, for which simulations indicate that shock velocities should exceed $\sim$25\,\kms\ \citep[e.g.,][]{gusdorf2008a,gusdorf2008b}. The lack of correlation between the observed spatial extent of gas-phase SiO and \HHCO\ suggests that it is not the same shocks responsible for both molecules; other aspects, such as local density and temperature, must also play a role in determining the morphology of the \HHCO\ emission map. The morphology of SiO emission, particularly on the southeast side of source~A, outlines a more collimated, redshifted outflow (jet) structure than the maps of \HHCO\ and \HthCN\ do (Fig.~\ref{fig:velocitymaps}). 

The blueshifted structure of \CtfS\ is seen deeper into the bridge than \HHCO\ and \HthCN, closer to the central axis of the bridge. Despite the overlap between the bridge and the \CtfS, the blueshifted velocities of the latter suggest that it is not part of the quiescent bridge. 
The redshifted emission component seen in \HHCO, \HthCN\ and \CtfS\ appears to be at the northernmost section of the redshifted outflow at position angle $\sim$270\degr\ \citep{yeh2008}. 
In summary, the two outflow-like emission shapes at position angles 335\degr\ and 285\degr\ (Fig.~\ref{fig:outflowcartoon}) are interpreted as two unrelated features. In our data, there is insufficient symmetry and alignment to assign them to the rotating structures around source~A at position angle (rotation axis) 326--360\degr\ \citep{favre2014a,oya2016}. We refer to Fig.~\ref{fig:outflowcartoon} for an illustration of the misalignment. 

The picture drawn in Fig.~\ref{fig:outflowcartoon} also addresses the coupling of small scales (60--1500~au), as seen in interferometric observations, with larger scales ($\sim$2000--30\,000~au) measured in single-dish observations. Firstly, we confirm the finding by \citet{yeh2008} that the east-west outflow pair at a few-hundred au scales is consistent with the much larger outflow lobes extending out to thousands of au in the same direction, likely both driven by component A in the IRAS\,16293 system. Secondly, the arcminute-scale outflow pair which is redshifted to the northeast and blueshifted to the southwest \citep{mizuno1990,stark2004} does not appear to have any counterpart at scales below $\sim$1000~au. This points to an outflow driven by IRAS\,16293A at some time in the past, but which has been quenched in recent times. Adopting an inclination angle of 65\degr\ 
for the northeast-southwest outflow lobes and a line-of-sight velocity offset of 10~\kms\ \citep{stark2004}, the absence of a northeast-southwest flow at scales $\lesssim$1000~au translates into a timeframe for the quenching of a few hundred years.

\begin{figure}
	\resizebox{\hsize}{!}{\includegraphics{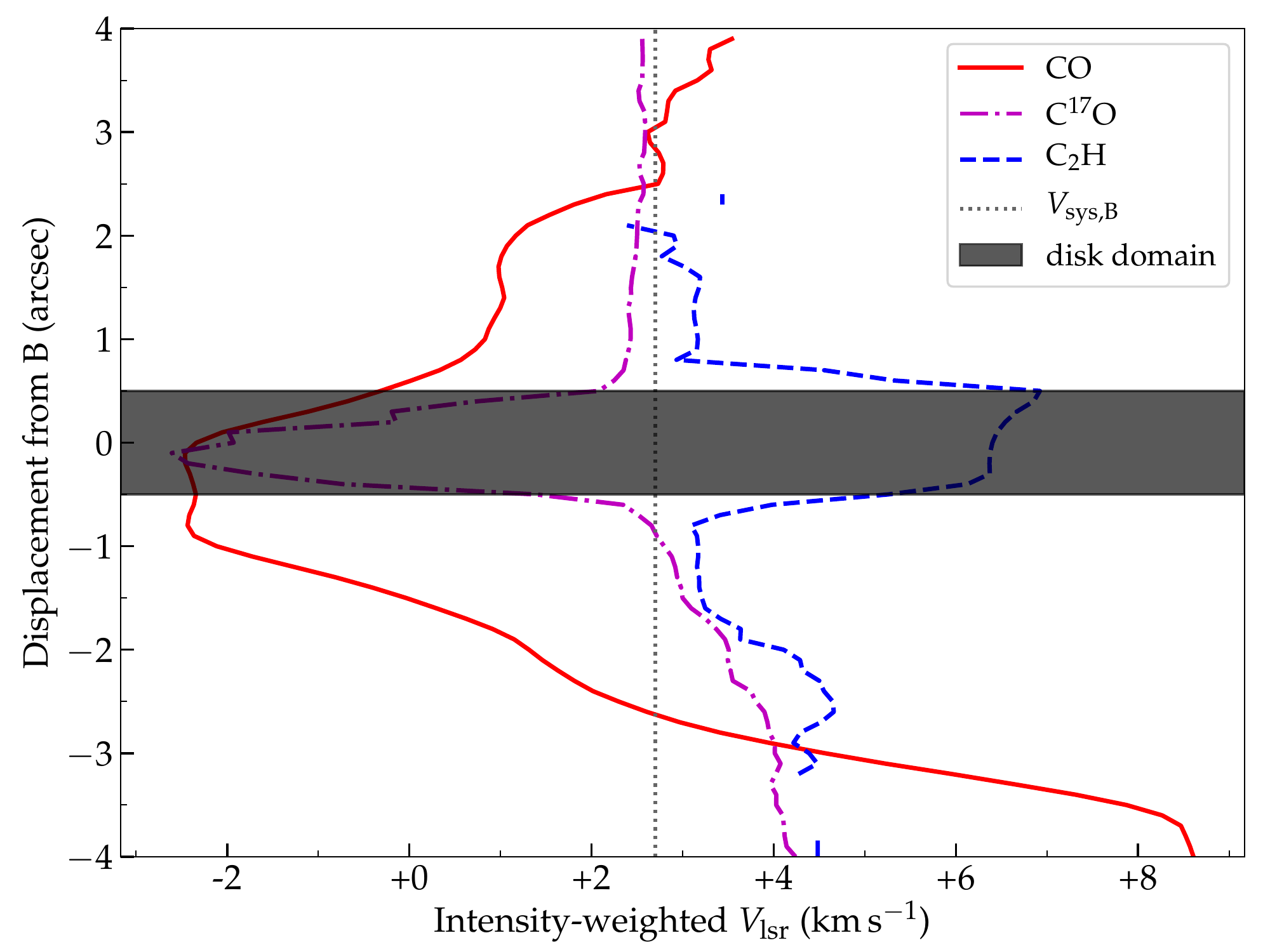}}
	\caption{Position-velocity diagram along the slice indicated by the arrow in Fig.~\ref{fig:highvelocityCO}. Velocity is the intensity-weighted peak velocity, as in Fig.~\ref{fig:velocitymaps}. Position displacement is defined to be positive to the north of source~B. The `disk domain', inside which intensity-weighted velocity values are not meaningful, is marked with a partly transparent gray box. }
	\label{fig:pv_sourceB}
\end{figure}

Finally, the strongly blueshifted components apparent in CO~3--2 (Fig.~\ref{fig:highvelocityCO}) near source~B are interpreted using the position-velocity diagram in Fig.~\ref{fig:pv_sourceB}. We choose a south-north slice in order to cover the \CCH\ emission on either side of source~B as well as the $^{12}$CO blob south of source~B. In the diagram, displacement values below $-3$\arcsec\ correspond to the regime of the northern edge of the outflow cone emanating westward from source~A, where redshifted velocities are observed. All weighted velocity values for the main CO isotopologue within 3\arcsec\ on either side of source~B, however, are significantly blueshifted with respect to the systemic velocity of source~B (vertical dotted black line in Fig.~\ref{fig:pv_sourceB}). This must mean that surface layers on the front side (facing the observer), probed by optically thick CO emission, are moving away from the center of mass of protostellar source~B. Already noted in the \citeyear{yeh2008} publication by \citeauthor{yeh2008} (their position `b2'), the origin of this kinematic structure is still under debate. It may be a compact (and therefore young) outflow feature driven by source B \citep{loinard2013,oya2018}, or alternatively, a bow shock feature related to the northwest outflow driven by A \citep{kristensen2013a}. If driven by B, it remains to be explained why the symmetrical counterpart is hidden from sight, even at displacements 2--3\arcsec\ north of source~B, where dust optical depth would not be sufficient to absorb line emission from the background. If driven by source~A, the relatively modest velocity of the NW outflow axis, $\lesssim$5~\kms\ with respect to either source (\HthCN\ and SiO panels in Fig.~\ref{fig:velocitymaps}), has somehow been translated into much higher velocities for the dense material in the outskirts of source~B (up to 10~\kms\ with respect to the protostar, see dark blue contours in Fig.~\ref{fig:highvelocityCO}). One conceivable scenario is that a dense stream of gas emanating from source A impacts dense material around source~B, which deflects part of the gas such that its velocity vector becomes more aligned with the line of sight direction, which leads to higher line-of-sight velocity offsets. It may also be that some other, yet unidentified dynamical process is contributing. 
In the same region where $^{12}$CO is blueshifted, the bulk mass traced by optically thin \CstO\ has velocities consistent with the systemic velocity of source~B. In the disk domain, masked in Fig.~\ref{fig:pv_sourceB} with a box spanning a width identical to the diameter of the masks in Fig.~\ref{fig:velocitymaps}, even \CstO\ is affected by the high optical depth of line and continuum photons. The C$_2$H emission, which also stretches north-south across source~B (Figs.~\ref{fig:velocitymaps}, \ref{fig:CCHchannelmaps}), shows a velocity trend in Fig.~\ref{fig:pv_sourceB} overlapping that of \CstO. These two species show a modest velocity gradient across source~B,: \vlsr\ is +1.8 to +2.7~\kms\ north of source~B, and +2.6 to +4.1~\kms\ south of B, whichever physical component it traces. 

Two pockets of SiO 8--7 emission are seen $\sim$1\arcsec\ north and south of source~B, at position angle 15\degr\ east of north (Fig.~\ref{fig:SiOchannelmaps}), not overlapping with the \CCH\ filament (Sect.~\ref{sec:CCHfilament}) at position angle $-15$\degr. Line of sight velocities of SiO in these pockets are shifted toward both blue and red sides of the systemic velocity, which makes their association to either outflow or inflow motions ambiguous. The much weaker, but optically thin emission from $^{29}$SiO (Fig.~\ref{fig:29SiOchannelmaps}) shows the same morphology, again with mixed blue and red velocities. The morphology and kinematics of SiO 8--7 in our map is consistent with that observed in SiO \mbox{7--6} by \citet{oya2018}, who interpret this structure as a signature of a pole-on pair of outflows.

\section{Discussion} 
\label{sec:discussion} 

To highlight the different physical and dynamical components studied in this work, the molecular gas observed between and around the binary protostellar system IRAS\,16293 at 60--1500~au scales is divided into three distinct domains: (i) dense and hot ($>$100~K) gas in the disk or disk-like regions around sources~A and B; (ii) more tenuous and colder gas residing in the dust bridge between the protostars; and (iii) kinematically active gas within or on the borders of outflow lobes driven by source~A. Domain (i) is not examined in this work, but its kinematics and temperature structure was extensively studied, using a different set of molecular tracers, by \citet{oya2016} for source~A and by \citet{oya2018} for source B. Domains (ii) and (iii) are discussed in Sects.~\ref{sec:bridgedomain} and \ref{sec:outflowdomain}. A structure seen in \CCH, seemingly unrelated to outflow or infall dynamics, is addressed in Sect.~\ref{sec:CCHfilament}. Finally, the relative evolutionary stage of the two sources in IRAS\,16293 is addressed in Sect.~\ref{sec:agedifference}. 

\subsection{The quiescent bridge}
\label{sec:bridgedomain}

The arc-like bridge structure between protostars A and B is made up of material which has a moderate density (\pow{4}{4}\,\pccm -- $\sim$\pow{3}{7}\,\pccm). Evidence for the density range is provided by the observation that the only molecular species clearly tracing the morphology of the dust arc is \CstO, while all tracers of higher density (see Table~\ref{t:selectedlines}) follow outflow structures and have line-of-sight velocity structures that deviate significantly from that of \CstO. This interpretation is supported when our observations are compared with three-dimensional models from \citetalias{jacobsen2018}, yielding a qualitative match in morphology of the observed bridge structure (Sect.~\ref{sec:bridgemodel}). 
To finally distinguish between the effects of gas temperature and gas density, observation would be needed of the distribution of a similarly high density tracer ($n_\mathrm{crit}>10^6$\,\pccm) which at the same time has a low upper level energy (\Eup $<$ 30~K). For example, the $J$=3--2 or $J$=2--1 transition of \HthCN\ both satisfy these conditions. 

We hypothesize that the bridge is a remnant substructure of the circumbinary envelope \citep{schoeier2002} or a filamentary core, from which both protostellar sources have formed in the past. With its current mass budget and temperature conditions, the IRAS\,16293 bridge filament is stable against further gravitational collapse (Sect.~\ref{sec:bridgefragmentation}). 
In addition, the bridge arc is kinematically quiescent, lying at a flat \vlsr\ within 0.5~\kms\ of the systemic velocities of both protostars (Fig.~\ref{fig:velocitymaps}, top left). This straightforward observational fact rules out a scenario in which the bridge arc is one segment of a large, circumbinary disk (or torus) in which both protostars would be embedded. In such a scenario, the bridge gas would have shown a line-of-sight velocity gradient following the kinematic signatures of sources A ($+$3.1~\kms) and B ($+$2.7~\kms), that is, a blue-to-red gradient in the southeast-to-northwest direction. This would only comply with the observed flat velocity distribution if the `disk' would rotate entirely in the plane of the sky, which is inconsistent with the line of sight velocity difference of the two protostars. Compared to other, tighter protostellar binary systems such as GG~Tau \citep{dutrey2014,dutrey2016} and IRS~43 \citep{brinch2016}, the bridge that we observe between the components of IRAS\,16293 lacks a symmetric complement to close a full circumbinary disk or torus, and its velocity structure is inconsistent with disk-like rotation. 
 If there is any velocity gradient across the bridge, it is in the transverse direction rather than along the length of the arc. 

Although the 600~au separation between the two components in IRAS\,16293 is somewhere in the mid-field between close and wide binaries (see~Sect.~\ref{sec:intro}), we conclude that the formation of protostars A and B has probably occurred through turbulent fragmentation. The reason is that the competing scenario, disk fragmentation \citep{adams1989}, is unlikely due to: (i) the lack of evidence for a remnant of a circumbinary disk, as discussed in detail in this work (see above); and (ii) the stark misalignment between the two disk-like structures, face-on for source~B \citep{jorgensen2016,oya2018}, and roughly edge-on for source~A \citep{pineda2012,girart2014}. 

Like IRAS\,16293, the more evolved Class~I binary protostellar system IRAS 04191$+$1523, with two components separated by 860~au, also shows a bridge of gas connecting the two protostellar sources \citep{lee_j-e2017}. Based on the velocity structure of the \CeiO\ bridge, \citet{lee_j-e2017} conclude that the two sources and the bridge are a substructure of the same natal envelope, which has given rise to protostar formation through turbulent fragmentation. The object L1521F, hosting only a very young, low-luminosity source \citep{bourke2006}, has also revealed a filamentary arc of gas (\HCOplus) when studied at high resolution with ALMA \citep{tokuda2014}. In this case, the filament itself is not detected in dust continuum emission; it connects to a compact millimeter continuum component at its western-most terminus, whereas protostars have not yet formed on its eastern end. The dense core \mbox{Barnard 5}, with at least one embedded protostar, also exhibits filamentary structure within which three additional, possibly prestellar condensations have formed \citep{pineda2015}. The physical origin of the filaments in both L1521F and in \mbox{Barnard 5} is unclear, but their long axis dimensions of $\gtrsim$1000\,au make a disk origin seem unlikely. 
In contrast, the Class 0 L1448\,IRS3B triple protostellar system hosts a heavily curved connecting filament seen in \CeiO, and its kinematics are consistent with being a remnant of a 254~au diameter disk that is hypothesized to have spawned all three cores through fragmentation of a disk or the innermost envelope \citep{tobin2016b}. A connecting bridge was also found in the triple system SR24, with a $\sim$700~au size, by \citet{fernandez-lopez2017}, who suggest that it, too, may be interpreted as a remnant of a disk spiral arm. 
In each of these case studies, the bridging filaments appear to have a relation to the nursery of the protostellar sources, either in the form of a protostellar disk undergoing fragmentation or a substructure of a larger-scale envelope. IRAS\,16293 appears to belong to the latter category. 

\subsection{Outflow signatures}
\label{sec:outflowdomain}

The SiO gas surrounding IRAS\,16293B at projected radii up to 250~au ($J$=8--7, Sect.~\ref{sec:kinematics} of this work; $J$=7--6, \citealt{oya2018}) is of mixed blue- and red-shifted velocity up to $\pm$3\,\kms\ from the systemic velocity of source~B. The presence of SiO in the gas phase is often used as a tracer of outflow shocks \citep[e.g.,][]{flower1996,caselli1997,gusdorf2008a,gusdorf2008b}. While the overlapping blueshifted and redshifted SiO emission is suggestive of a geometry with approaching and receding outflows superposed in projection \citep{oya2018}, the relatively low velocities would not be expected to lead to grain erosion. Correcting the observed velocity for projection effects has a marginal effect, since the \vlsr\ offset is necessarily almost completely parallel with the true velocity vectors in case of a face-on disk launching an outflow along the line of sight. However, in the scenario that the SiO pockets near source~B are due to impact of the northwest outflow from source~A, the currently observed velocities of 5--10\,\kms\ may not lead to sufficiently strong shocks. Before silicate material from the grain cores is released into the gas phase, relative velocities should exceed $\sim$15\kms\ \citep[e.g.,][]{caselli1997,gusdorf2008a}. This leaves neither the pole-on outflow from source~B nor the impact of the outflow from A onto source~B as a fully viable scenario. 

The morphology of two molecular lines that are sensitive to densities above $10^7$~\pccm, o-\HHCO~5$_{1,5}$--4$_{1,4}$ and \HthCN\ \mbox{4--3} (Table~\ref{t:selectedlines}), are compared with a three-dimensional model in Sect.~\ref{sec:bridgemodel}. In the observed maps (Fig.~\ref{fig:velocitymaps}), these high-density tracers do not coincide with the axis of the dust/\CstO\ bridge. The modeled structure (\citetalias{jacobsen2018}; Sect.~\ref{sec:bridgemodel} of this work) does not produce emission in \HHCO\ or \HthCN\ from its bridge domain, consistent with the observations. Instead, the modeled emission is concentrated in the disk domains (and is extremely optically thick for o-\HHCO), with an extended hour-glass at positions above the disk plane of source~A, qualitatively consistent with enhanced pockets of emission seen in the observed maps southeast and northwest of the edge-on source~A. This is caused by self-absorption in the high-density midplane of the disk or disk-like structure, whereas its surface regions are less opaque, but still sufficiently dense and hot to excite the \HthCN\ into $J$=4. The absence of \HthCN\ in the bridge in the models is due to the chosen ice sublimation temperature of 100~K, which is only exceeded in the innermost regions close to the protostars. 

The high-density molecular gas observed in emission near the bridge filament is interpreted as being the effect of two outflows at position angles 315\degr\ and 270\degr, impacting patches of higher density in the ambient envelope. The northernmost of the two outflows is blueshifted, and is posited to be material on the southwestern edge of the bridge, swept up by the blueshifted side of the northwest-southeast outflow pair (Sect.~\ref{sec:kinematics}, Fig.~\ref{fig:outflowcartoon}). The southern, redshifted component at position angle 285\degr, coincides with the northern border of the redshifted western outflow cone (\citealt{yeh2008}; Fig.~\ref{fig:highvelocityCO} in this work). The same molecules are not observed at the southern edge of the western outflow; we speculate that this is due to a density contrast in the ambient medium, with higher density at the northern side, toward the bridge filament. \HHCO\ in particular is known to sublimate efficiently from dust grain surfaces not just by thermal desorption, but also by photodesorption such as occurs in classical photodissociation regions \citep{vanderwiel2009,guzman2011}. Chemical models of outflow-envelope interface regions \citep{drozdovskaya2015} also show enhanced \HHCO\ abundance, at early times. It is therefore natural to expect enhanced gas-phase \HHCO\ abundances in directly irradiated walls of outflow cavities in young protostellar systems such as IRAS\,16293. 

\subsection{Peak location for complex organic molecules}
\label{sec:sweetspot}

Searches for spectral signatures of relatively faint molecular line signals from new, complex (organic) molecules in the IRAS\,16293 system often focus on positions close to source~B, rather than source~A. The primary reason is the simpler, narrower spectral line profiles near source~B, owing to the difference in orientation of the disk-like structures surrounding sources~A and B, and the multitude of dynamically active outflow components associated with source~A (Sect.~\ref{sec:kinematics}, \ref{sec:outflowdomain}). With intrinsically narrower, symmetric line profiles, the disambiguation of nearby, partially overlapping lines is more straightforward. 
Specifically, the emission strength of lines of complex molecules is seen to peak in a region west-southwest of source~B at about 0.5\arcsec\ from its continuum peak \citep[e.g.,][]{baryshev2015,coutens2016,jorgensen2016,lykke2017,ligterink2017,calcutt2018a}. The offset can be explained by the effect of high optical depth of both line and continuum emission closer to the source center. The comparative lack of complex molecule emission on the north and east side of source~B may be related to the bridge filament connecting to source~B on that side (Sect.~\ref{sec:bridgedomain}), where physical conditions, such as temperature and dust optical depth, are likely different. In addition, excitation conditions and spectral line shapes on the east side of source~B may be affected by the northwest outflow from source~A impacting there (Sect.~\ref{sec:outflowdomain}). Finally, even to the south and west of source~B, dynamical effects do play a role in shaping molecular emission, as shown for example in the SiO morphology (Sect.~\ref{sec:results}, Fig.~\ref{fig:velocitymaps}). 



\subsection{The C$_\mathsf{2}$H filament} 
\label{sec:CCHfilament}

A narrow filament of material stretching across source~B in the south-north direction is visible in \CCH, and a similar, but slightly offset patch of high-velocity (\vlsr$<$$-6$~\kms) $^{12}$CO up to 1\farcs5 south of source~B. Compared with the \CCH\ emission observed in single-dish JCMT observations \citep{caux2011}, it is evident that the majority of \CCH\ flux is filtered out in our interferometric ALMA observations, much more than for spectral lines of other species. However, regardless of missing flux from a smooth, large-scale component, our ALMA \CCH\ map reveals compact substructures at scales $\sim$$10^2$-$10^3$~au. 
The velocity profiles of $^{12}$CO, \CstO\ and \CCH\ along this feature are shown in Fig.~\ref{fig:pv_sourceB}. \CCH\ has a rather flat velocity structure along the length of the filament, at \vlsr\ within $\sim$1~\kms\ of that of source~B itself, as already apparent from its intensity-weighted velocity map in Fig.~\ref{fig:velocitymaps}. The morphology of this feature, combined with its small line of sight velocity gradient, would be more consistent with a symmetric (outflow or infall) motion almost exactly in the plane of the sky, than with an outflow emanating perpendicularly from the face-on disk of source~B. Moreover, the observed spatial distribution of \CCH\ is not consistent with tracing one or multiple walls of outflow cavities carved out by any of the outflows in this system, such as has been observed in \CCH\ morphologies in other protostellar system at similar spatial scales, for example IRAS\,15398$-$3359 \citep{bjerkeli2016a} and VLA\,1623$-$2417 \citep{murillo2018}. 

The morphology of \CCH\ in IRAS\,16293 is also studied in \citet{murillo2018}, who find that it anti-correlates with that of cyclic C$_3$H$_2$. While non-overlapping distributions of these two chemically related species \citep{gerin2011} may seem surprising, there are also models and observations that indicate somewhat different physical conditions being needed for \CCH\ and $c$-C$_3$H$_2$ emission lines to occur \citep[e.g.,][]{cuadrado2015}. Lacking an obvious connection to any of the other features connected to source~A, the physical, dynamical and chemical origin of the C$_2$H emission remains an open question. 

\subsection{The relative evolutionary stage of sources~A and B}
\label{sec:agedifference}

Based on the conclusions drawn in this paper and accompanying PILS papers (\citetalias{jacobsen2018}; \citealt{calcutt2018a}), it remains unclear if sources A and B in the IRAS\,16293 system are coeval. Their rotation axes are not aligned and a formation scenario from a giant circumbinary disk is therefore unlikely (see Sect.~\ref{sec:bridgedomain}). Sources A and B differ in various aspects. Here we list those of relevance to the matter of evolutionary stage. 
($i$) Source~A is known to drive at least two sets of bipolar outflows, and an additional one in the past, whereas source~B shows signs of infall, but exhibits no unambiguous proof of outflow activity (\citealt{pineda2012}; \citealt{loinard2013}; \citealt{kristensen2013a}; Sect.~\ref{sec:kinematics} of this work). 
($ii$) \citetalias{jacobsen2018} find that the observed dust emission of the face-on disk of source~B can only be explained with a vertically extended disk or disk-like model (i.e., one that has not yet undergone significant settling), while the vertical structure of the disk of source~A is not constrained by their model. 
($iii$) Chemical differentiation is seen between both sources particularly when analyzing CN-bearing species. Such species have time dependent formation pathways \citep{garrod2017} with some, such as vinyl cyanide (C$_2$H$_3$CN), being formed later during the warm-up phase, as a result of the destruction of ethyl cyanide (C$_2$H$_5$CN). \citet{calcutt2018a} find vinyl cyanide to be at least nine times more abundant toward the B source. However, these authors stress that although this could indicate that source~A is less evolved, it could also imply that the warm-up timescales for source~A were very short, limiting the amount of vinyl cyanide that could be formed. 
Collecting the above information elements, ($i$) and ($ii$) are consistent with a view that source~B is at an earlier stage of evolution than source~A, whereas ($iii$) could be explained either way, appreciating that the temperature history of each source is not necessarily monotonous and smooth. We thus pose that source~B is at the same evolutionary stage or less evolved than source~A. 

If there is a difference in evolutionary stage between sources A and B, outflowing material from source~A may have provided a trigger for gravitational collapse to set in, some six hundred au (projected) further down its own natal filament. Alternatively, even if the initial collapse of source~B started spontaneously, the northwest outflow from source~A could now be feeding gas and dust originating in protostar~A onto the disk of source~B. Such a scenario would add a new element to the investigation of the different levels of various groups of chemical compounds found in sources~A and B \citep{bisschop2008,calcutt2018a}. 


\section{Summary and conclusions}
\label{sec:conclusions}

We have presented an analysis of the distribution and kinematics of molecular gas in the protostellar binary system IRAS\,16293 at scales of 60--2000~au, using ALMA-PILS observational data and three-dimensional line radiative transfer modeling. The selected tracers are 0.87~mm dust continuum, line transitions CO~3--2, \CstO~3--2, \HthCN~4--3, o-\HHCO~5$_{1,5}$--4$_{1,4}$, \HHthCO~5$_{1,5}$--4$_{1,4}$, \CtfS~7--6, SiO 8--7, $^{29}$SiO~8--7, and \CCH~4$_{7/2}$--3$_{5/2}$, together highlighting different aspect of the morphology and dynamics of the system.  
The main conclusions are summarized here. 
\begin{enumerate}
	\item The arc-shaped filament connecting the two protostellar sources is clearly seen in dust continuum and \CstO~3--2 emission. The kinematic pattern of \CstO\ indicates that it is quiescent, and consistent with the systemic velocities of both sources. Other molecular tracers considered in this work show a different spatial distribution, different kinematics, or both (Sect.~\ref{sec:results}). The gas density of the bridge filament is likely to be between \pow{4}{4}\,\pccm\ and $\sim$\pow{3}{7}\,\pccm\ (Sect.~\ref{sec:bridgemodel}). A three-dimensional radiative transfer model qualitatively matches the observed structure in dust and \CstO, but not those in \HthCN\ and \HHCO\ emission (Sect.~\ref{sec:bridgemodel}). 
	\item Using a simplified description of the bridge filament, balancing thermal pressure and gravity, the bridge is inferred to be stable against radial collapse. While a radially stable filament is a prerequisite for it to fragment along the direction of its axis, our interpretation of the observations of the bridge filament in its current state does not necessarily extend to earlier times when it was likely more massive and thus less stable (Sect.~\ref{sec:bridgefragmentation}).
	\item Given the physical nature of the bridge filament, we pose that it is a remnant substructure of a filamentary circumbinary envelope that has undergone turbulent fragmentation to form both protostellar sources (Sect.~\ref{sec:bridgedomain}). The possibility that the bridge is (a remnant of) a circumbinary disk is ruled out (Sect.~\ref{sec:bridgedomain}).  
	\item The western side of the east-west outflow pair is clearly traced in CO~3--2 in our observations, shown to be accelerating with distance from the launching source~A. In addition, it is seen to change from a collimated to a more wide-angle structure at a projected distance of $\sim$350~au from source~A, possibly related to a density and pressure gradient of one of the more or less cylindrical structures observed in this system. The launching point of the outflow may be connected to a H$_2$O maser shock spot $\sim$120~au west of source~A (Sect.~\ref{sec:kinematics}). 
	\item Molecular lines of (isotopologues of) \HHCO, HCN, SiO, and CS trace outflow motions or the impact of outflows onto more quiescent components. Dynamic structures near source~B might be due to nearly pole-on outflow from source~B itself, or impact of the northwest outflow from source~A onto dense material in source~B (Sect.~\ref{sec:kinematics}). In our observations, we find no counterpart for the large scale (thousands of au) northeast-southwest outflow pair, indicating that its launching engine must have been quenched a few hundred years ago (Sect.~\ref{sec:kinematics}). 
	\item A striking, previously unknown structure stretches straight across source~B at position angle $\sim$15\degr. It is only seen in \CCH\ and is kinematically flat, which rules out an origin in outflow activity of source~B (Sect.~\ref{sec:CCHfilament}). 
	\item Combining evidence from this work and other studies, we suggest that IRAS\,16293B is not more evolved than its sister source~A (Sect.~\ref{sec:agedifference}). 
	\item When used as a target to search for complex (organic) molecules, as is often done, the physical and dynamical complexity of the IRAS\,16293 protostellar binary system should be taken into account. While searches for complex molecules generally avoid the broad line regions near source~A, even the often-used positions on or near the blue `$\times$' mark in Fig.~\ref{fig:velocitymaps}, 0.5\arcsec\ offset from source~B, may exhibit spectral line shapes with evidence of line-of-sight motions, which will manifest differently in different molecules (Sect.~\ref{sec:sweetspot}). 
	
\end{enumerate} 

By abandoning spherically symmetric model descriptions of the envelopes of IRAS\,16293 \citep[e.g.,][]{jorgensen2002,schoeier2002,crimier2010b} and other protostellar systems, \citetalias{jacobsen2018} have taken a significant leap by developing a three-dimensional density model that includes an envelope hosting two individual energy sources, A and B, disk-like structures around each, an interbinary bridge filament, and a self-consistently derived temperature structure of the ensemble. While the radiative transfer scheme initially included dust and CO isotopologues, it was further expanded in this work to include two additional molecular species, \HHCO\ and \HthCN. However, there are still steps to be made to make such a model more realistic. For example, chemistry, freezing out and sublimation of molecules could be included, beyond the current implementation using gas-phase abundance jumps at certain threshold temperatures. The current model also lacks various dynamic components such as the outflows, and the rotating, disk-like structure around source~A, which may need to be expanded to larger radii than the 150~au used in \citetalias{jacobsen2018}. In addition, future observational work could answer currently open questions about, for example, (i) the evolutionary state of source~B, (ii) the debated small-scale multiplicity of source~A, or (iii) the connection between the relatively young, small-scale outflow signatures (a few hundred to a few thousands~au) observed in interferometric observations such as in this paper, and the older outflows seen at distances of tens of thousands of au from the protostellar sources. 
Considering the already known level of complexity of the system, and the additional aspects yet to be uncovered, adding all these elements to a (self-consistent) physical model would further increase the number of degrees of freedom, which may result in challenges when attempting to sensibly constrain the model parameters. 

The protostellar stages of this and other binary systems should be studied further in order to understand how interactions with binary companions (through direct illumination or even mass transfer through an outflow) affect the evolution of the inner tens to hundreds of au of such systems. From this work and the literature so far, it appears that IRAS\,16293 represents a particular stage in the evolution of binary star systems, a phase which many other protostars may go through at one point in their evolution. IRAS\,16293 may not be particularly special; it is thus rather appropriate to use this system to study chemical and physical conditions in a representative analog of the young Solar system. 

\begin{acknowledgements} 
We thank the anonymous referee for thoughtful comments that helped improve the quality of this paper.
MHDvdW, SKJ, JKJ, LEK and HC acknowledge support from the European Research Council (ERC) under the European Union's Horizon 2020 research and innovation program (grant agreement No.~646908) through ERC Consolidator Grant ``S4F'', as well as from a Lundbeck Foundation Group Leader Fellowship. 
PB acknowledges support from the Swedish Research Council through contract 637-2013-472. Research at the Centre for Star and Planet Formation is funded by the Danish National Research Foundation. 
The postdoctoral grant of AC is funded by the ERC Starting Grant 3DICE (grant agreement 336474). 
MND acknowledges the financial support of the Center for Space and Habitability (CSH) Fellowship and the IAU Gruber Foundation Fellowship. 
CF acknowledges support from the Italian Ministry of Education, Universities and Research, through the grant project SIR (RBSI14ZRHR). 
This paper makes use of the following ALMA data: ADS/JAO.ALMA\#2013.1.00278.S. ALMA is a partnership of ESO (representing its member states), NSF (USA) and NINS (Japan), together with NRC (Canada), NSC and ASIAA (Taiwan), and KASI (Republic of Korea), in co-operation with the Republic of Chile. The Joint ALMA Observatory is operated by ESO, AUI/NRAO and NAOJ. 
This research has made use of NASA's Astrophysics Data System Bibliographic Services. 
This research has made use of APLpy, an open-source plotting package for Python \citep{aplpy2012}; Astropy, a community-developed core Python package for Astronomy (Astropy Collaboration \citeyear{astropy2013}), the matplotlib plotting library \citep{matplotlib2007}, and the CASA data processing package \citep{mcmullin2007}. 
\end{acknowledgements}

\bibliographystyle{aa}  
\bibliography{../../literature/allreferences}

\clearpage

\begin{appendix}

\section{Line contamination} 
\label{sec:contamination} 

Since the PILS data set, covering 329.15--362.90~GHz, has a line density of about one line every 3~MHz \citep{jorgensen2016}, it is important to consider the effect of contamination of other species contributing to the observed signal of a selected tracer. The potential contributions from species and transitions other than those listed in Table~2 are visualized in the spectral profiles in Figs.~\ref{fig:C17Ospectrum}--\ref{fig:CCHspectrum} below. 

In general, while various contaminating contributions are identified in the spectrum extracted toward the position one beam west-southwest of source~B, this signal is exclusively due to relatively complex molecular species, which are known to have spatial distributions confined to the regions in the central 1--2\arcsec\ around sources~A and/or B \citep[e.g.,][]{baryshev2015,lykke2017,ligterink2017}, domains explicitly not under study in the current work. The only exception is an SO$_2$ transition near \HthCN~4--3 profiles. This and other cases that require a more detailed description are addressed in the following paragraphs. 

\begin{figure}[!b]
	\resizebox{\hsize}{!}{\includegraphics[angle=0]{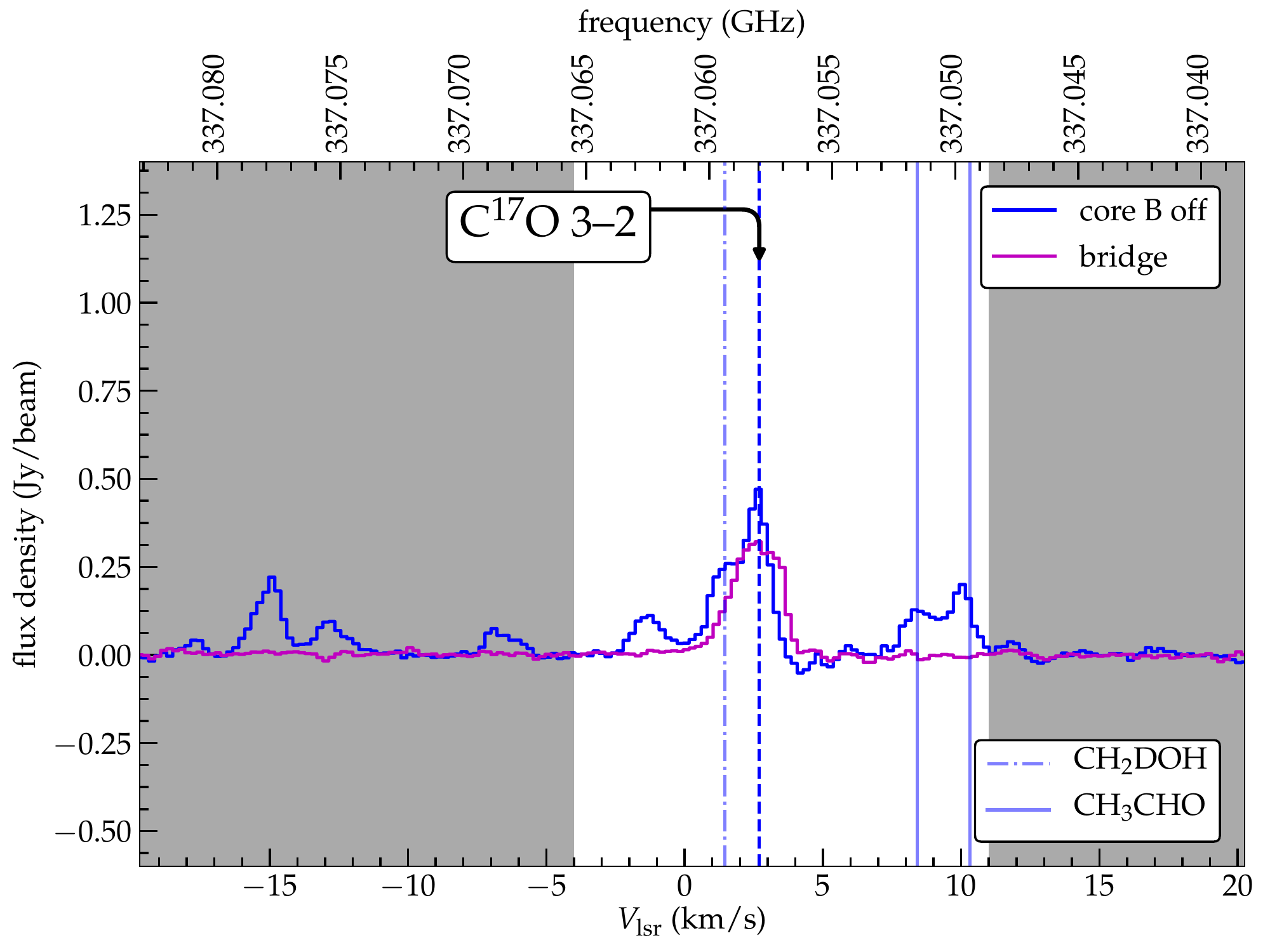}}
	\caption{
	Spectral profiles of \CstO~3--2 at two spatial positions: one beam west-southwest of source~B (blue `$\times$' mark in Fig.~\ref{fig:velocitymaps}) and within the bridge filament (magenta `$+$' mark in the same figure). The frequency of the species of interest is marked by the blue, vertical dashed line; other lines identified in the frequency range are marked with dotted, dash-dotted, or solid lines. All rest frequencies are shifted by $+2.7$~\kms\ to account for the systemic velocity of source~B. The section of the velocity axis with white background denotes the maximum range over which intensities are integrated.  
	}
	\label{fig:C17Ospectrum}
\end{figure}

\begin{figure}[!h]
	\resizebox{\hsize}{!}{\includegraphics[angle=0]{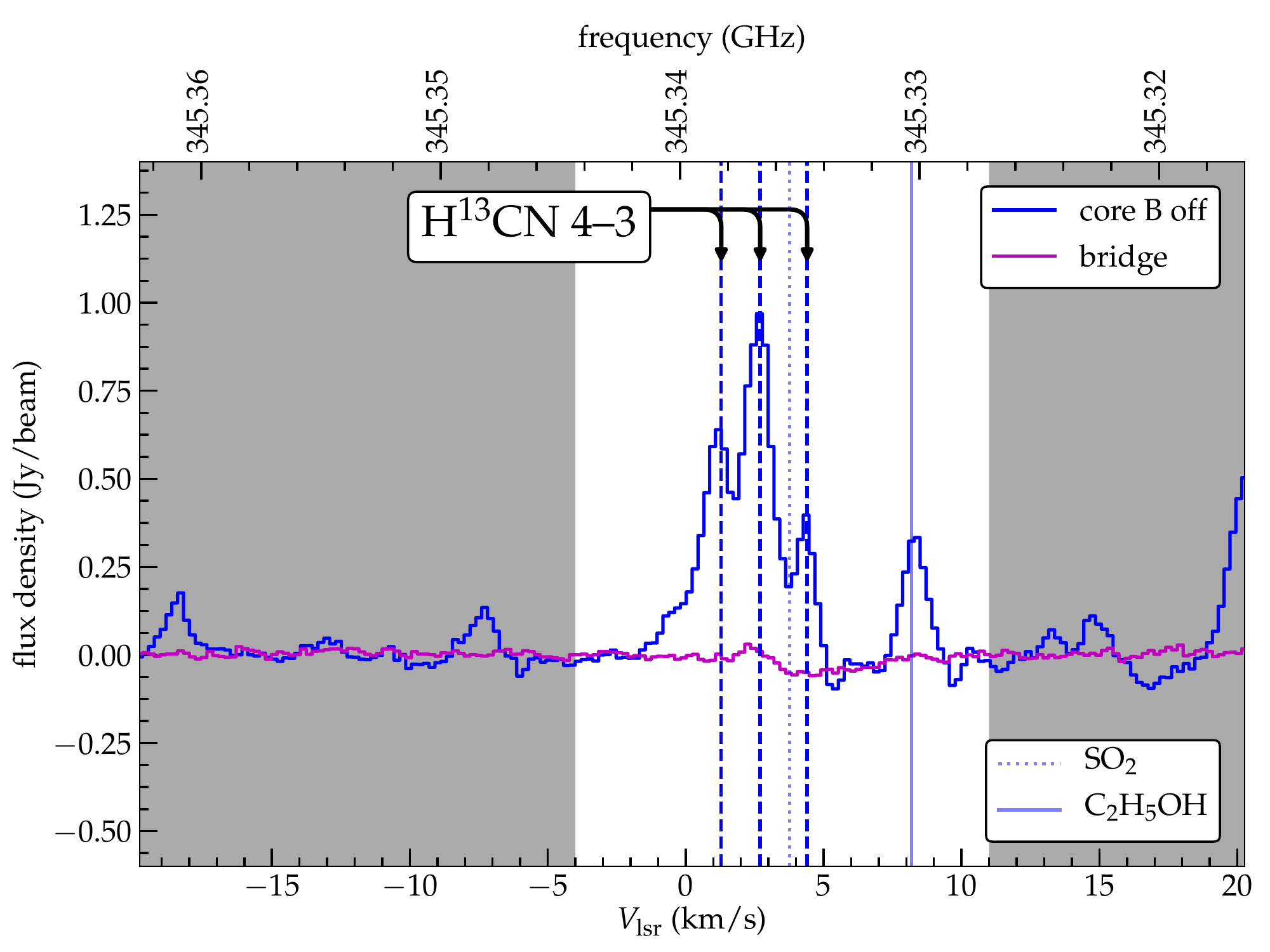}}
	\caption{
	Same as Fig.~\ref{fig:C17Ospectrum}, but for \HthCN. Its $J$=4--3 transition has three hyperfine components marked with the black arrows, in order of increasing frequency (right to left): $F$=4--4, several transitions with $\Delta F$=1, and $F$=3--3. 
	}
	\label{fig:H13CNspectrum}
\end{figure}

\paragraph{\CstO.} 
The hyperfine splitting for the \CstO~3--2 rotational transition is only 0.5~MHz. We therefore indicate the rest frequency of this transition with a single vertical dashed line in Fig.~\ref{fig:C17Ospectrum}, derived from the mean frequency of the contributing hyperfine components with Einstein $A$ coefficients \mbox{$>$$10^{-6}$ s$^{-1}$}.

\paragraph{\HthCN~4--3.}
Of the three separated hyperfine components of this rotational transition \citep{cazzoli2005}, the central one ($\Delta F$=1) is taken as the reference for the velocity axis. 
There is potential contamination of the \mbox{\HthCN~4--3} signal by the \SOtwo\ 13$_{2,12}$--12$_{1,11}$ transition (345.3385377 GHz), with a modest upper level energy of 92\,K, at only 1.2~MHz from the rest frequency of \HthCN. At the position 0.5\arcsec\ southwest of source~B (`$\times$' mark in Fig.~\ref{fig:velocitymaps}), preferred for identifying spectral signatures of rare, complex molecules in previous PILS papers \citep[e.g.,][]{coutens2016,lykke2017,ligterink2017}, lines are narrow and spectrally resolved, and we do not see evidence of \SOtwo\ contribution to the \HthCN\ total intensity beyond a few percent. At other positions, for example in the western outflow from source~A or close to source~A itself, contamination cannot be ruled out. 

\paragraph{$^{29}$SiO~8--7.} 
The signal of the $^{29}$SiO~8--7 itself is weak and somewhat blueshifted ($\sim$1\,\kms) toward the position 0.5\arcsec\ offset from source~B (Fig.~\ref{fig:29SiOspectrum}). We are confident, however, that the signal ascribed to $^{29}$SiO in the channel maps (Fig.~\ref{fig:29SiOchannelmaps}) and velocity map (Fig.~\ref{fig:velocitymaps}), because the morphology of the redshifted $^{29}$SiO channel maps follows that of the brightest patches of SiO (Fig.~\ref{fig:SiOchannelmaps}) and the spectral line shape toward those positions (for example, the extension southeast of source~A) is single-peaked. 

\paragraph{\CCH~4$_{7/2}$--3$_{5/2}$.} 
For the \CCH\ $N_J$=4$_{7/2}$--3$_{5/2}$ transition, there are two hyperfine structure components ($F$=4--3 and 3--2), separated by 1.4~MHz. Such a small separation means that they lie within 1.2~\kms\ of one another in the spectral dimension, but are resolved in our 0.2~\kms\ spectral resolution data cubes, which complicates the interpretation of its velocity structure. In this work, the reference frequency (zero-point for \vlsr) is taken as the mid-point between the two hyperfine structure lines, $F$=4--3 at  349.39928 GHz and $F$=3--2 at 349.40067 GHz \citep{padovani2009}.  
In addition, most of the redshifted (>$+5$~\kms) emission in the velocity map of \CCH\ (Fig.~\ref{fig:velocitymaps}, Fig.~\ref{fig:CCHchannelmaps}) is due to the bright CH$_3$CN~19$_3$--18$_3$ transition at 349.39329~GHz. For this reason, we have adopted a narrower integration for \CCH\ than for the other lines (cf.~Fig.~\ref{fig:CCHspectrum} vs.~Fig.~\ref{fig:C17Ospectrum}--\ref{fig:29SiOspectrum}). In positions near sources~A and B, the CH$_3$CN line also contaminates channels closer to the systemic velocity of \CCH.

\begin{figure}[!h]
	\resizebox{\hsize}{!}{\includegraphics[angle=0]{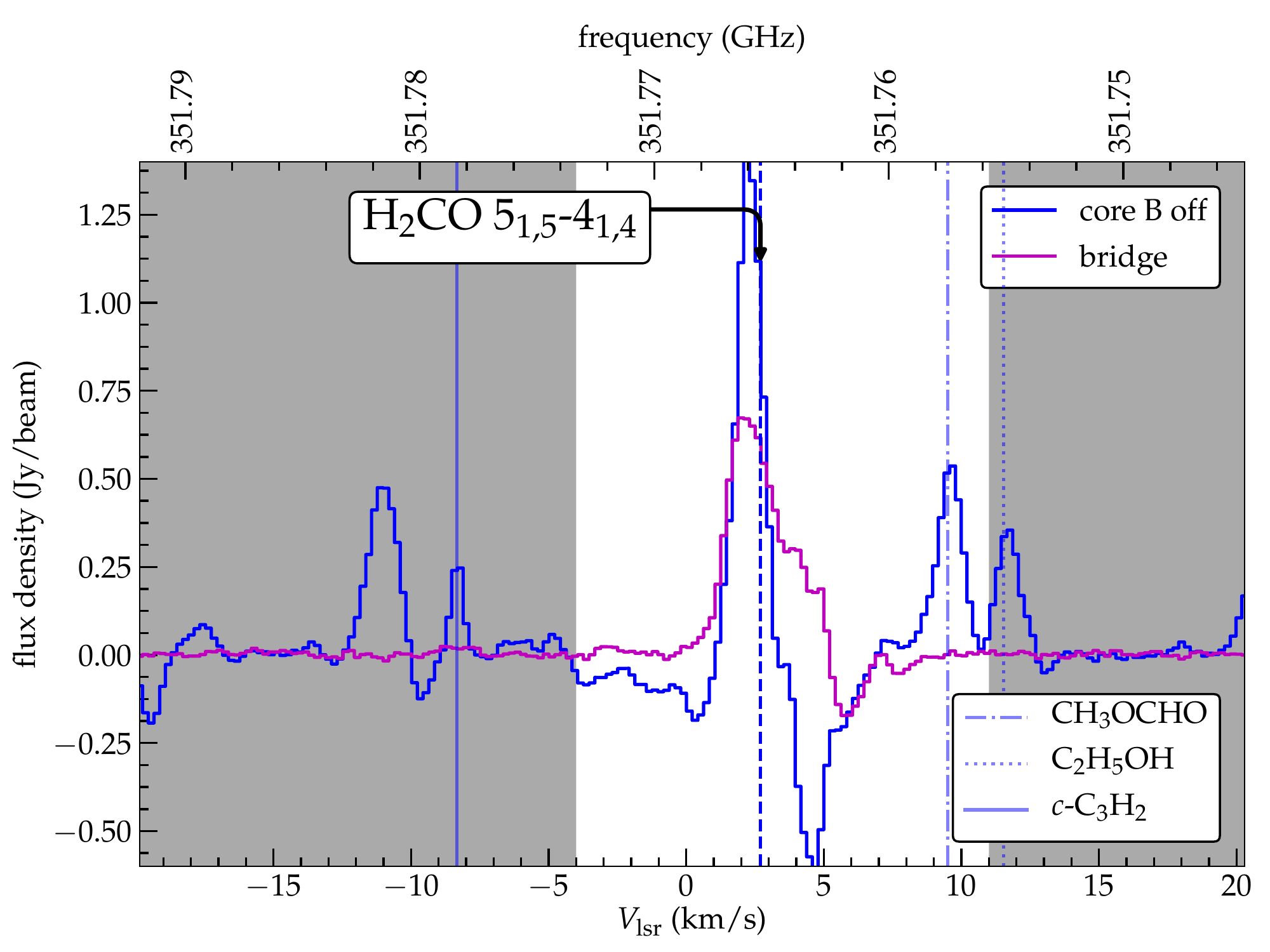}}
	\caption{Same as Fig.~\ref{fig:C17Ospectrum}, but for \HHCO.}
	\label{fig:H2COspectrum}
\end{figure}

\begin{figure}[!h]
	\resizebox{\hsize}{!}{\includegraphics[angle=0]{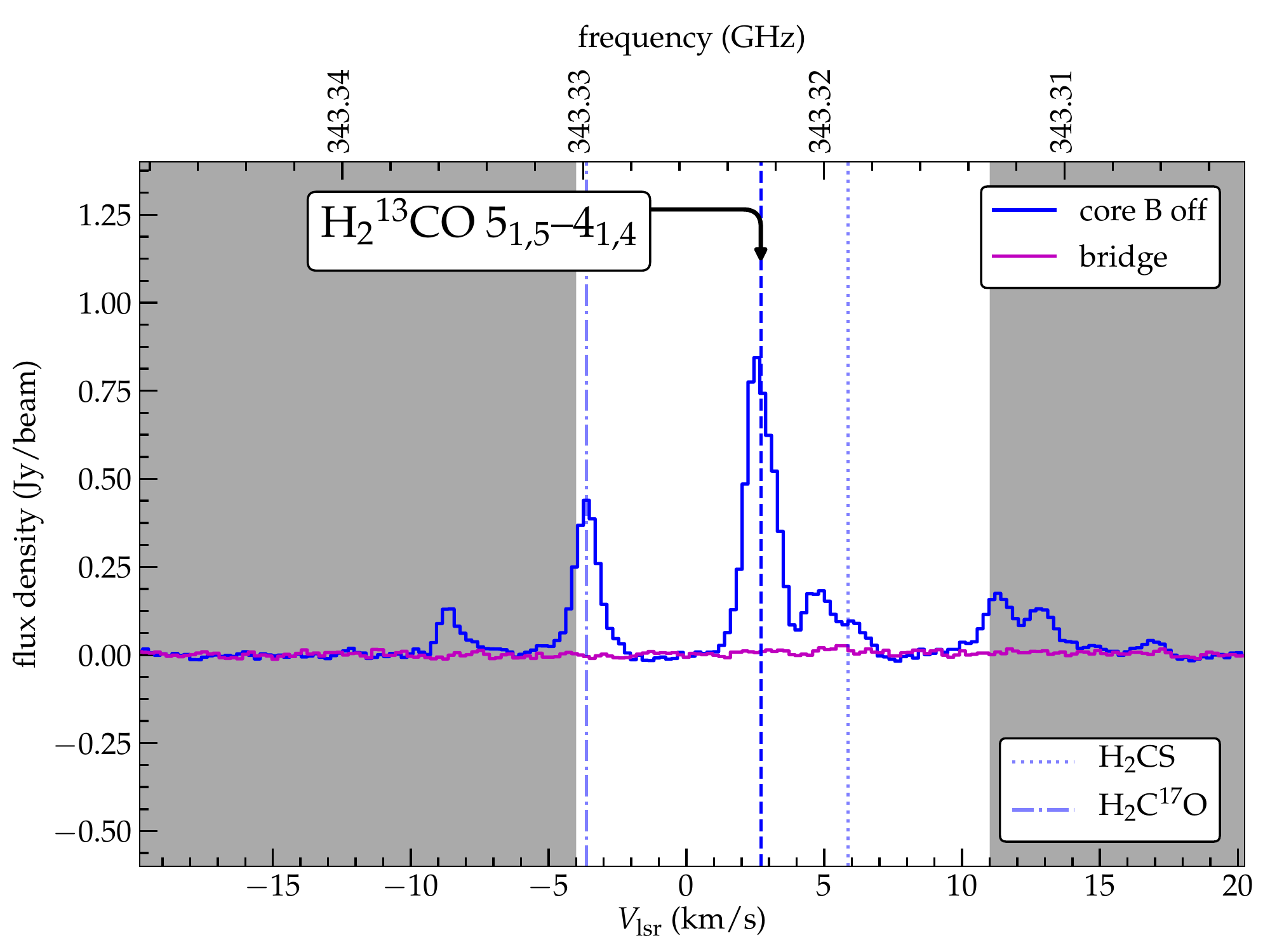}}
	\caption{Same as Fig.~\ref{fig:C17Ospectrum}, but for \HHthCO.}
	\label{fig:H213COspectrum}
\end{figure}

\begin{figure}[htb]
	\resizebox{\hsize}{!}{\includegraphics[angle=0]{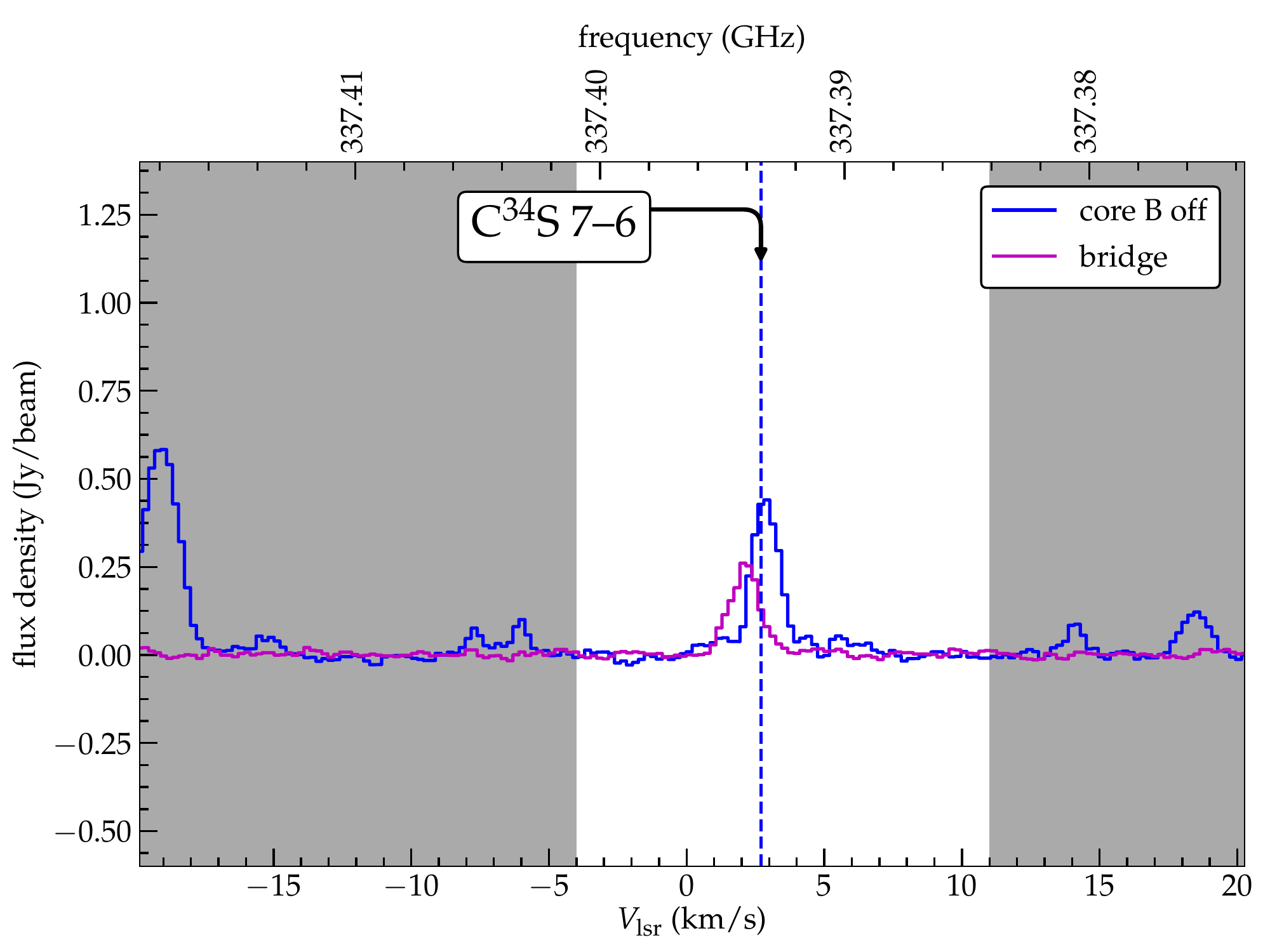}}
	\caption{Same as Fig.~\ref{fig:C17Ospectrum}, but for \CtfS.}
	\label{fig:C34Sspectrum}
\end{figure}

\begin{figure}[htb]
	\resizebox{\hsize}{!}{\includegraphics[angle=0]{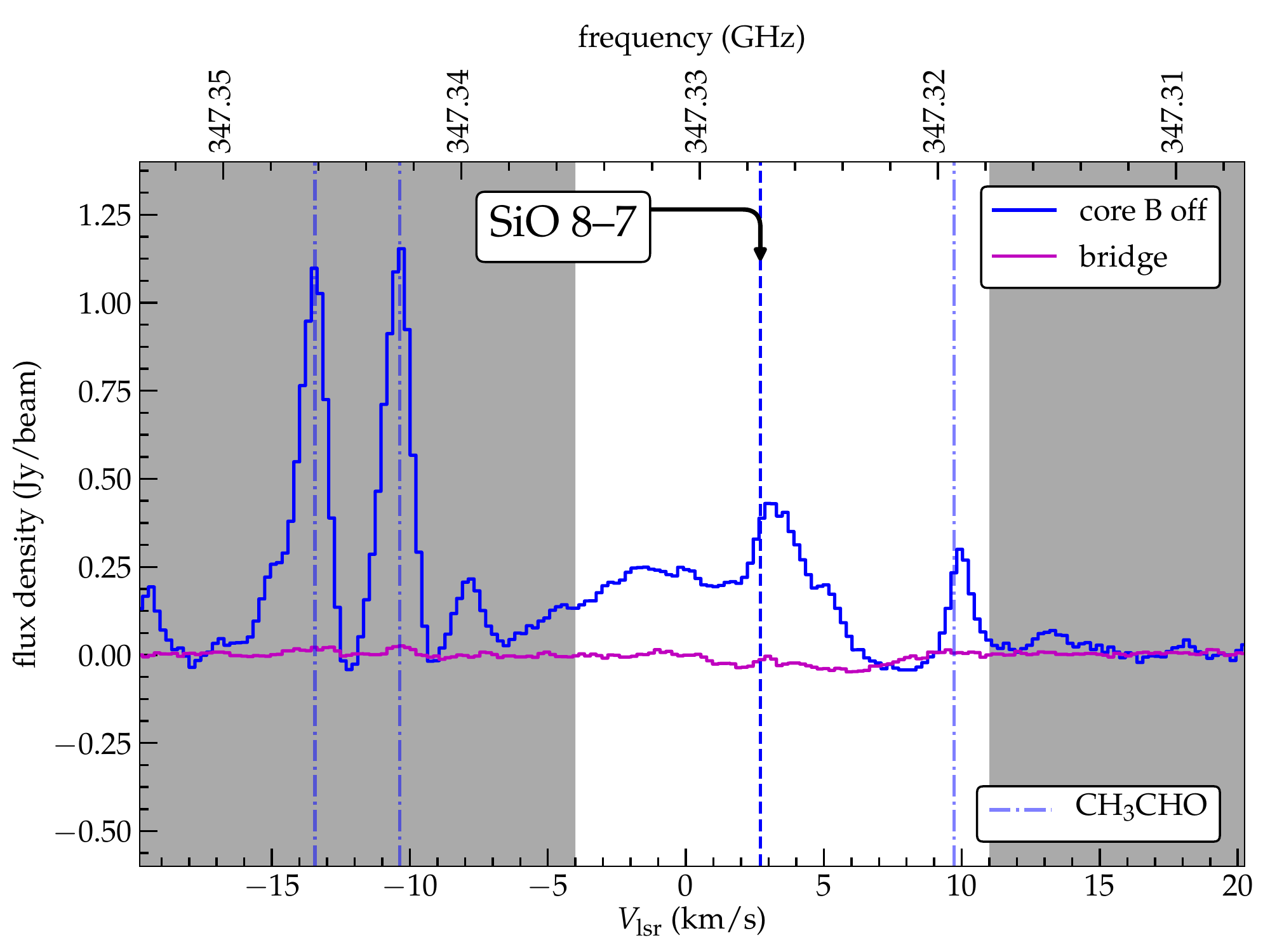}}
	\caption{Same as Fig.~\ref{fig:C17Ospectrum}, but for SiO.}
	\label{fig:SiOspectrum}
\end{figure}	

\begin{figure}[htb]
	\resizebox{\hsize}{!}{\includegraphics[angle=0]{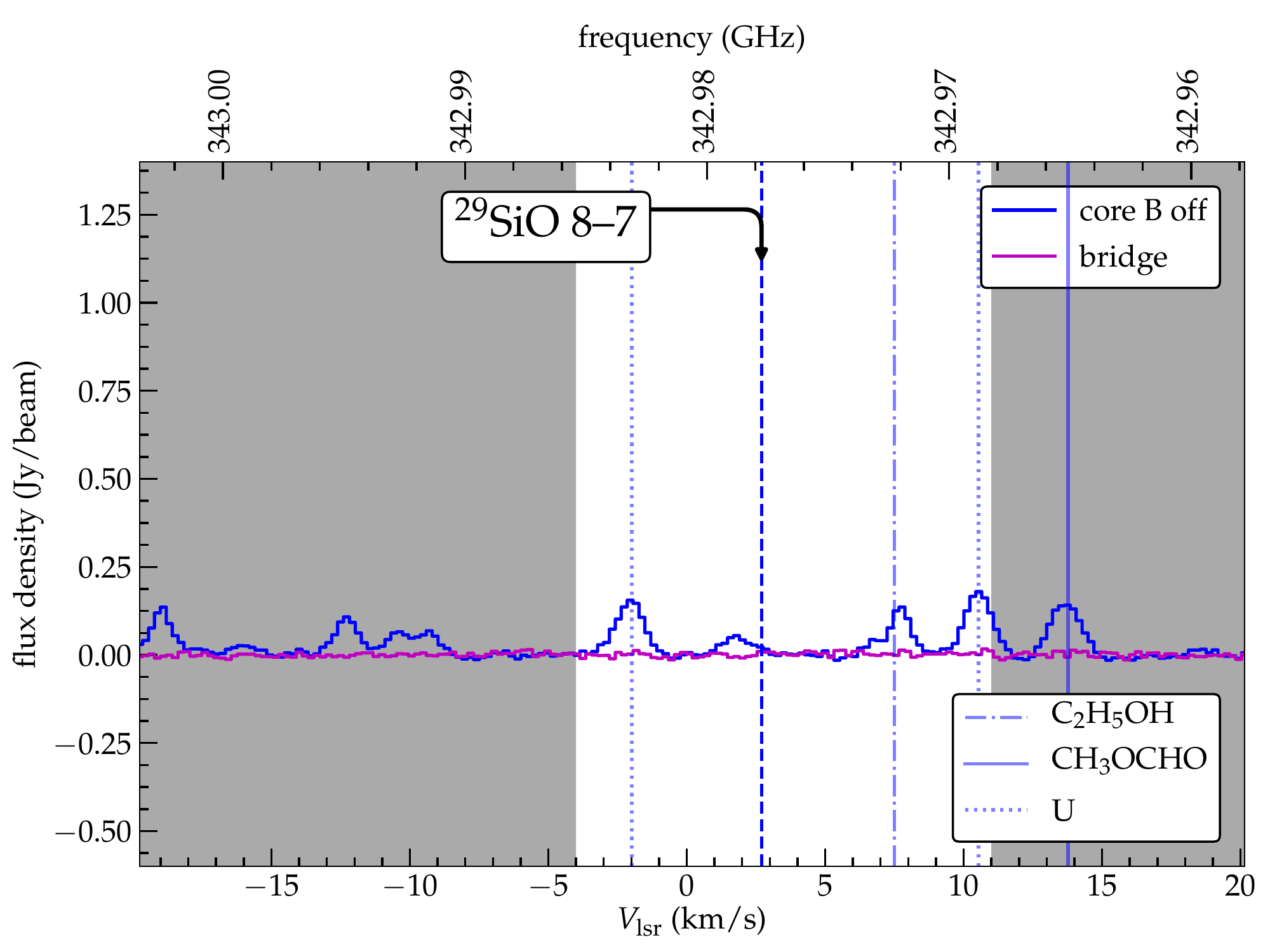}}
	\caption{Same as Fig.~\ref{fig:C17Ospectrum}, but for $^{29}$SiO. }
	\label{fig:29SiOspectrum}
\end{figure}	

\begin{figure}[htb]
	\resizebox{\hsize}{!}{\includegraphics[angle=0]{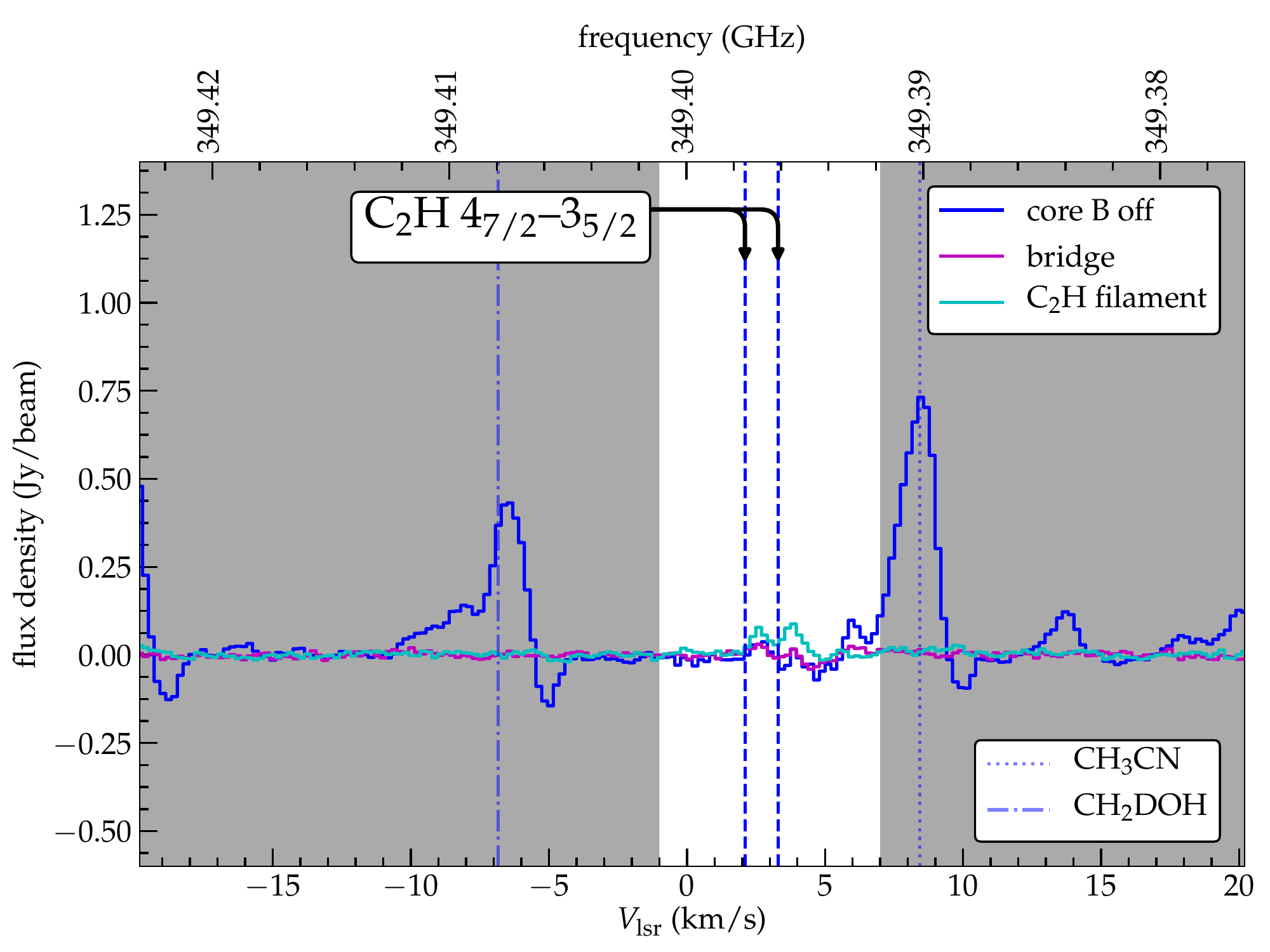}}
	\caption{Same as Fig.~\ref{fig:C17Ospectrum}, but for \CCH.}
	\label{fig:CCHspectrum}
\end{figure}

\section{Channel map figures}
\label{sec:channelmapfigures}

Channel maps of the molecular lines listed in Table~\ref{t:selectedlines} are shown in Fig.~\ref{fig:C17Ochannelmaps}--\ref{fig:CCHchannelmaps}.

\begin{figure*}
	\resizebox{\hsize}{!}{\includegraphics[angle=0]{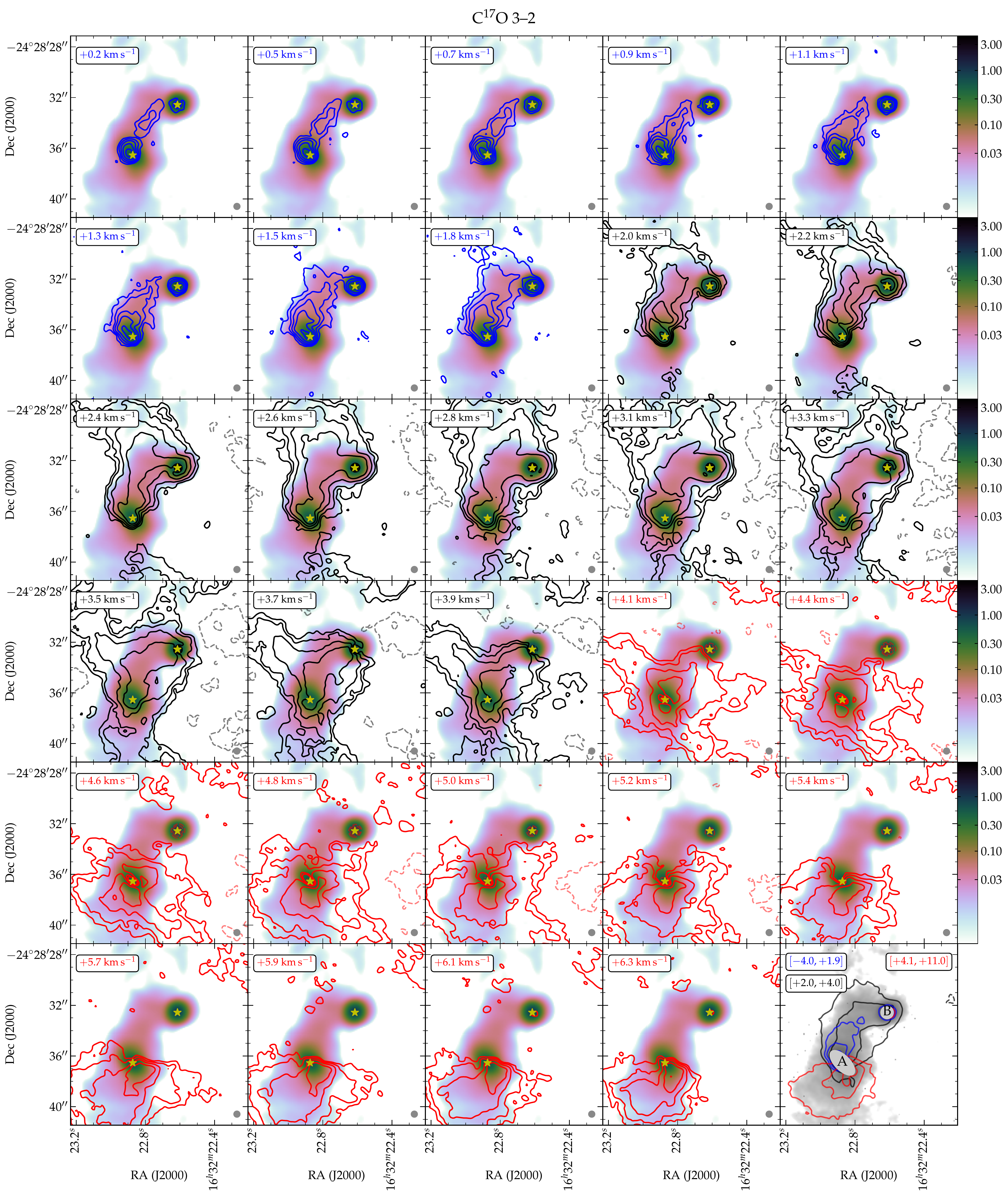}}
	\caption{\CstO~3--2 channel maps with contours representing logarithmically spaced line intensity in each channel. The first positive contour (solid blue, black, or red) is at 0.03 \Jyperbeam, the first negative contour (lighter, dashed) is at $-0.06$ \Jyperbeam, with every next contour level spaced by a factor of two. The bottom right panel displays three integrated velocity ranges in colored contours: all `blue' channels, [$-4.0$, $+2.0$$\rangle$ \kms; all `systemic velocity' channels, [+2.0, +4.0] \kms; and all `red' channels, $\langle$+4.0, +11.0] \kms. For each of these three, the first contour level is at 0.2 \Jyperbeam\,\kms\ and levels increase by factors of two, and equivalently for negative contours displayed in dashed line format. Gray scale in bottom right panel (logarithmic stretch from 0.002 to 2.0 \Jyperbeam), color scale in all other panels (logarithmic stretch indicated in color bar): 0.87~mm continuum. Star symbols mark the continuum peak locations of protostars A and B (see Table~\ref{t:peakcoordinates}). }
	\label{fig:C17Ochannelmaps}
\end{figure*}

\begin{figure*}
	\resizebox{\hsize}{!}{\includegraphics[angle=0]{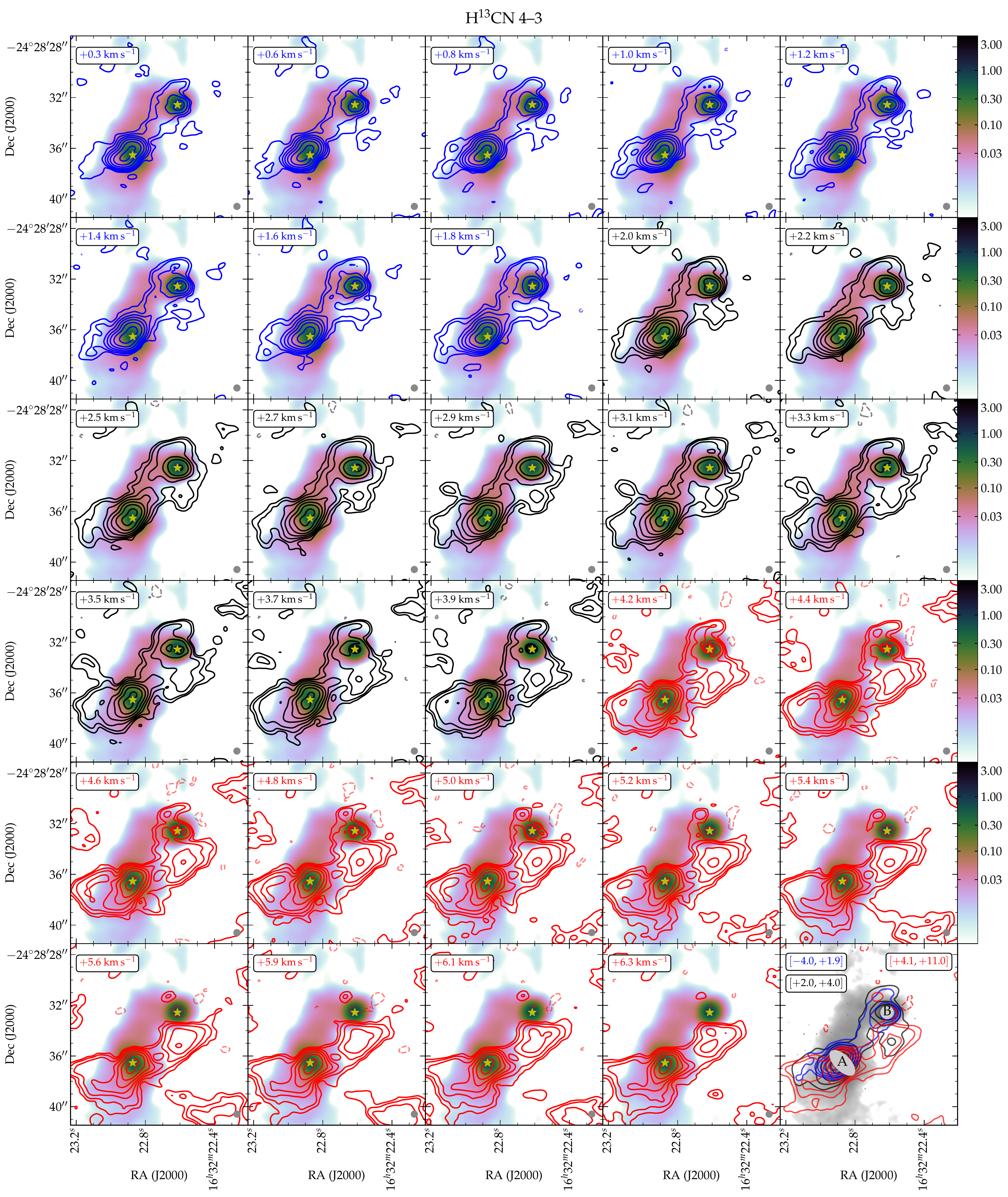}}
	\caption{\HthCN~4--3 channel maps. As Fig.~\ref{fig:C17Ochannelmaps}. First positive (negative) contour level in individual channel maps: 0.03 ($-0.06$) \Jyperbeam; first contour levels in the bottom right integrated ranges panel: 0.2 \Jyperbeam\,\kms.}
	\label{fig:H13CNchannelmaps}
\end{figure*}

\begin{figure*}
	\resizebox{\hsize}{!}{\includegraphics[angle=0]{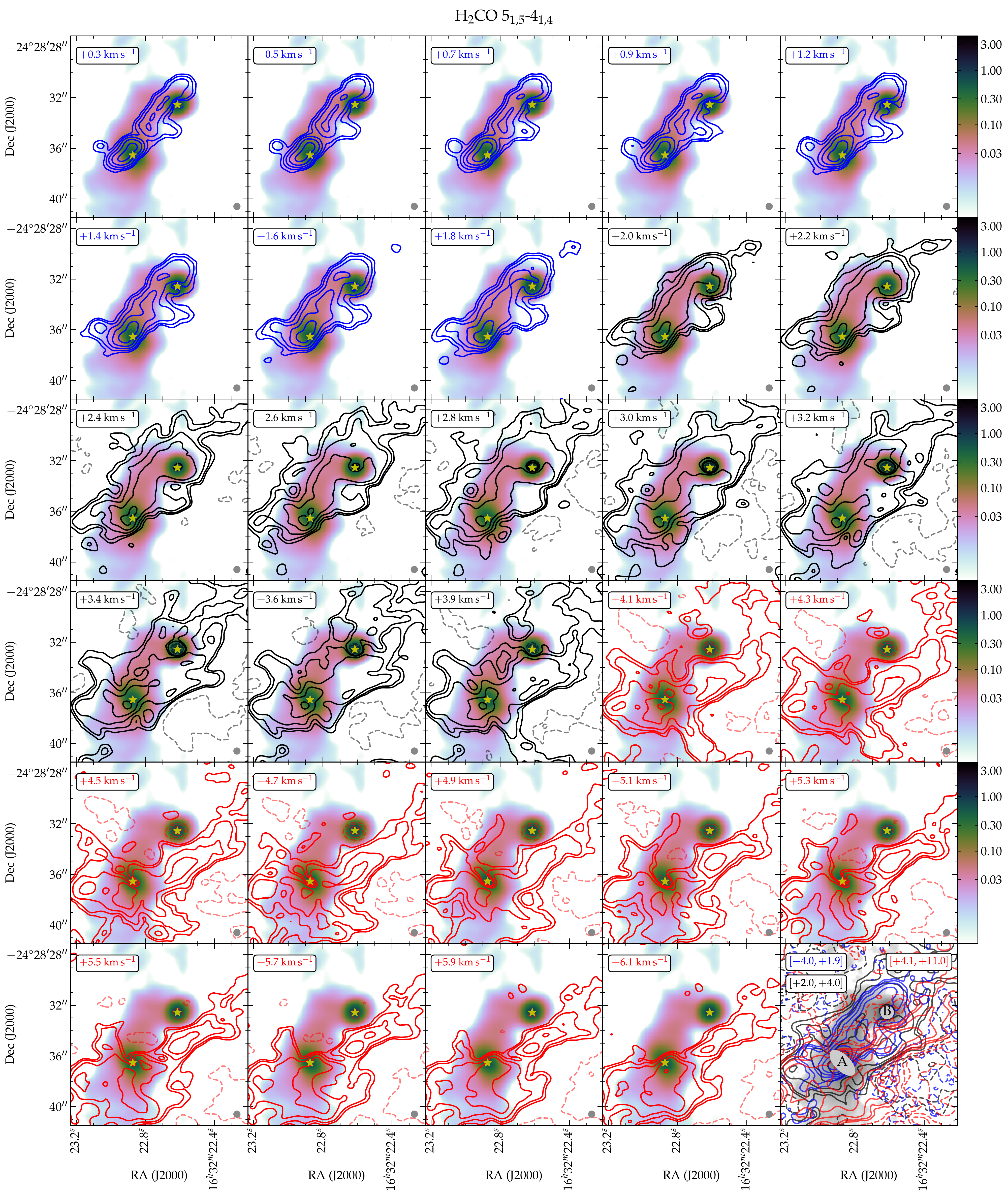}}
	\caption{\HHCO~5$_{1,5}$--4$_{1,4}$ channel maps. As Fig.~\ref{fig:C17Ochannelmaps}. First positive (negative) contour level in individual channel maps: 0.12 ($-0.24$) \Jyperbeam; first contour levels in the bottom right integrated ranges panel: 0.2 \Jyperbeam\,\kms.}
	\label{fig:H2COchannelmaps}
\end{figure*}

\begin{figure*}
	\resizebox{\hsize}{!}{\includegraphics[angle=0]{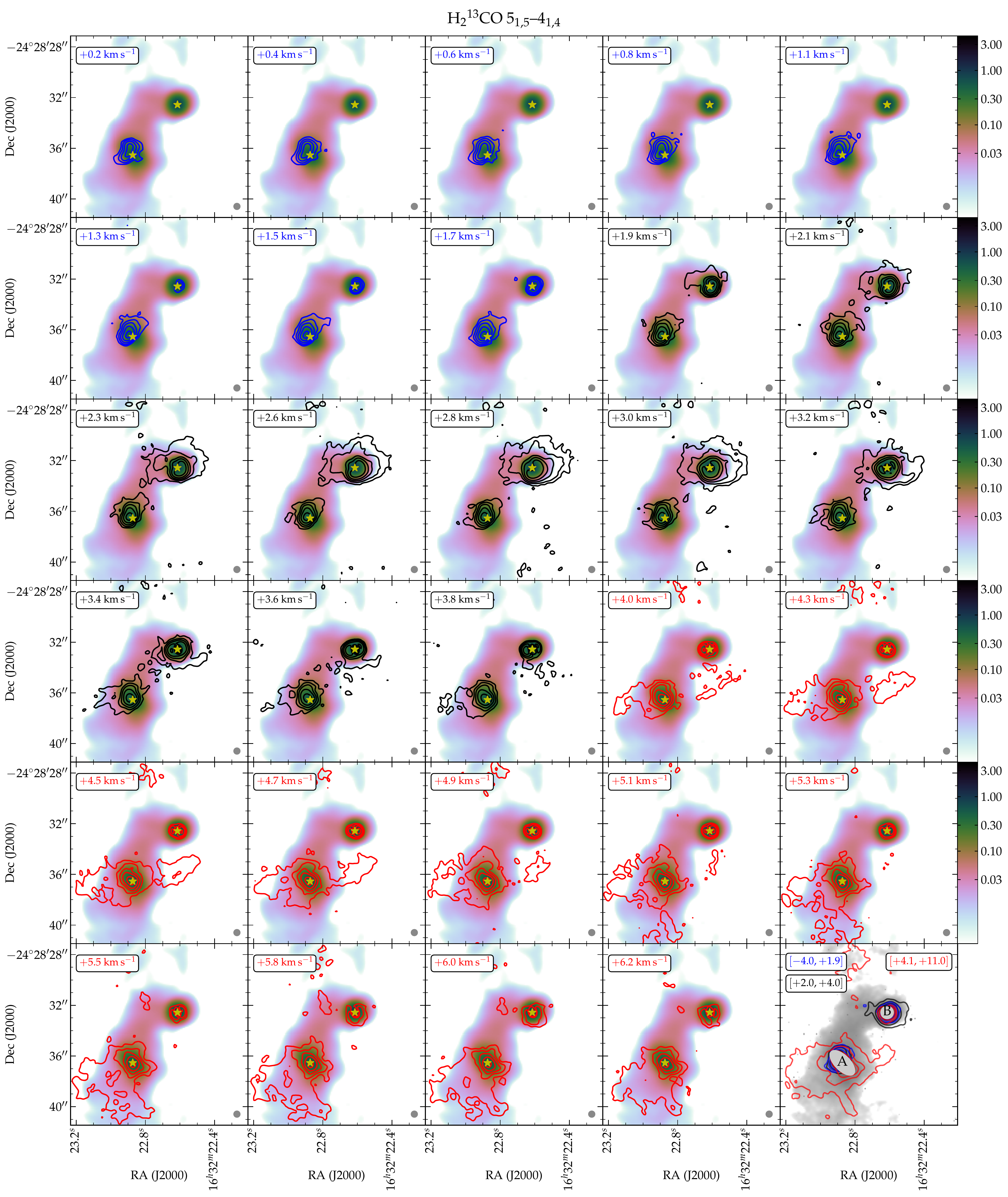}}
	\caption{\HHthCO~5$_{1,5}$--4$_{1,4}$ channel maps. As Fig.~\ref{fig:C17Ochannelmaps}. First positive (negative) contour level in individual channel maps: 0.03 ($-0.06$) \Jyperbeam; first contour levels in the bottom right integrated ranges panel: 0.1 \Jyperbeam\,\kms. }
	\label{fig:H2[13]COchannelmaps}
\end{figure*}

\begin{figure*}
	\resizebox{\hsize}{!}{\includegraphics[angle=0]{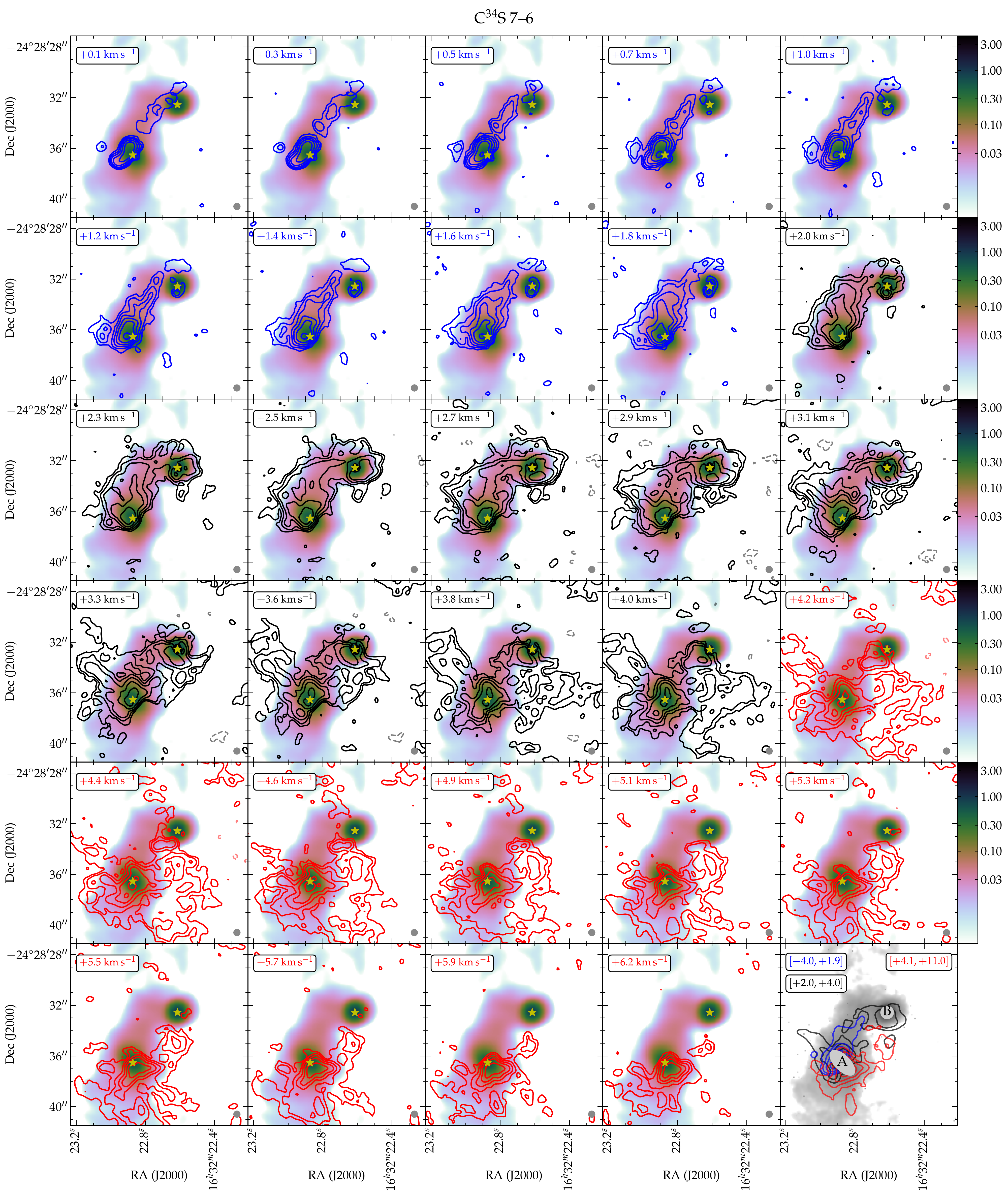}}
	\caption{
	\CtfS\ channel maps. As Fig.~\ref{fig:C17Ochannelmaps}. First positive (negative) contour level in individual channel maps: 0.03 ($-0.06$) \Jyperbeam; first contour levels in the bottom right integrated ranges panel: 0.2 \Jyperbeam\,\kms. 
	 }
	\label{fig:C34Schannelmaps}
\end{figure*}

\begin{figure*}
	\resizebox{\hsize}{!}{\includegraphics[angle=0]{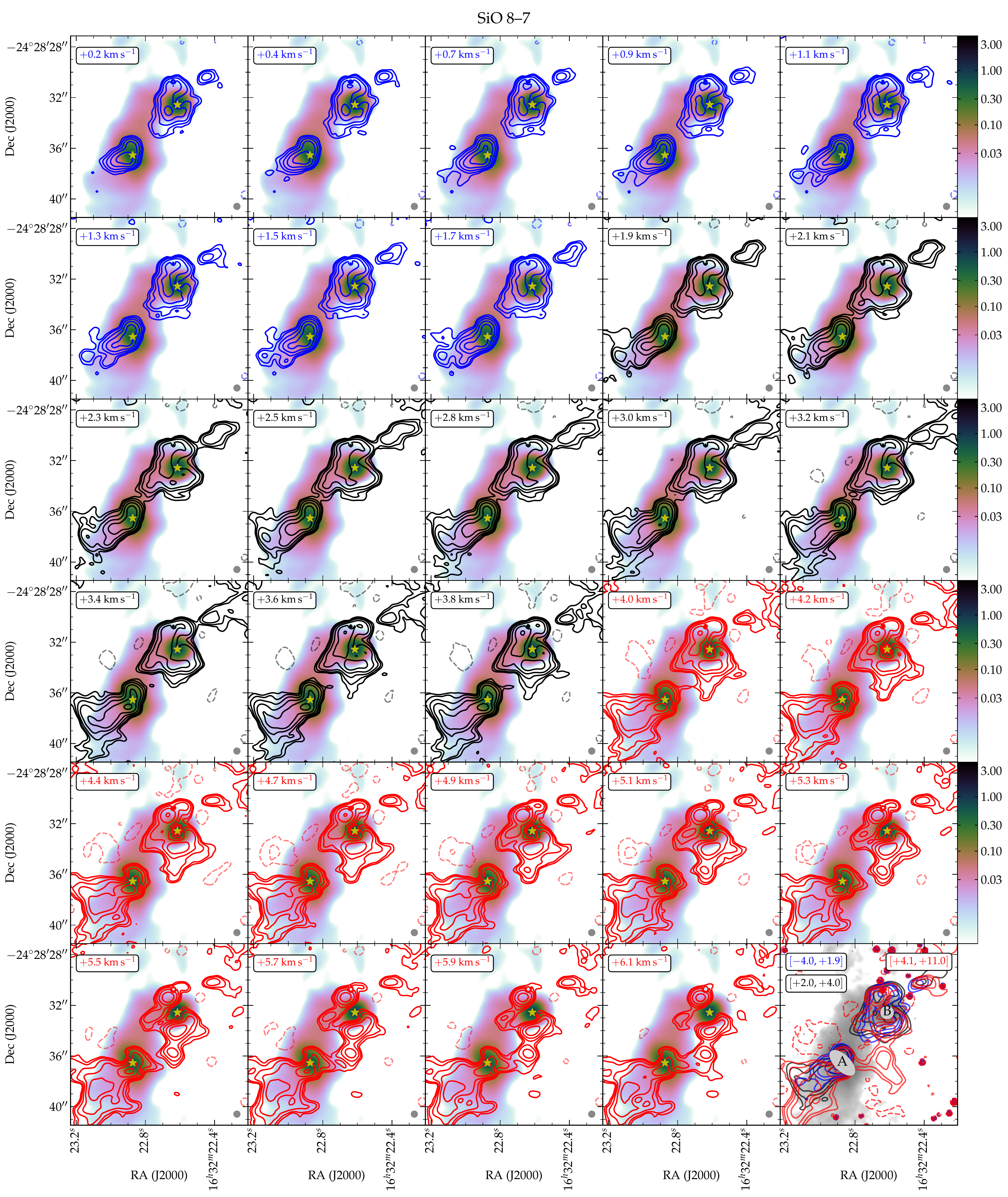}}
	\caption{
	SiO (main isotopologue) 8--7 channel maps. 
	As Fig.~\ref{fig:C17Ochannelmaps}. First positive (negative) contour level in individual channel maps: 0.03 ($-0.06$) \Jyperbeam; first contour levels in the bottom right integrated ranges panel: 0.2 \Jyperbeam\,\kms. 	
	 }
	\label{fig:SiOchannelmaps}
\end{figure*}

\begin{figure*}
	\resizebox{\hsize}{!}{\includegraphics[angle=0]{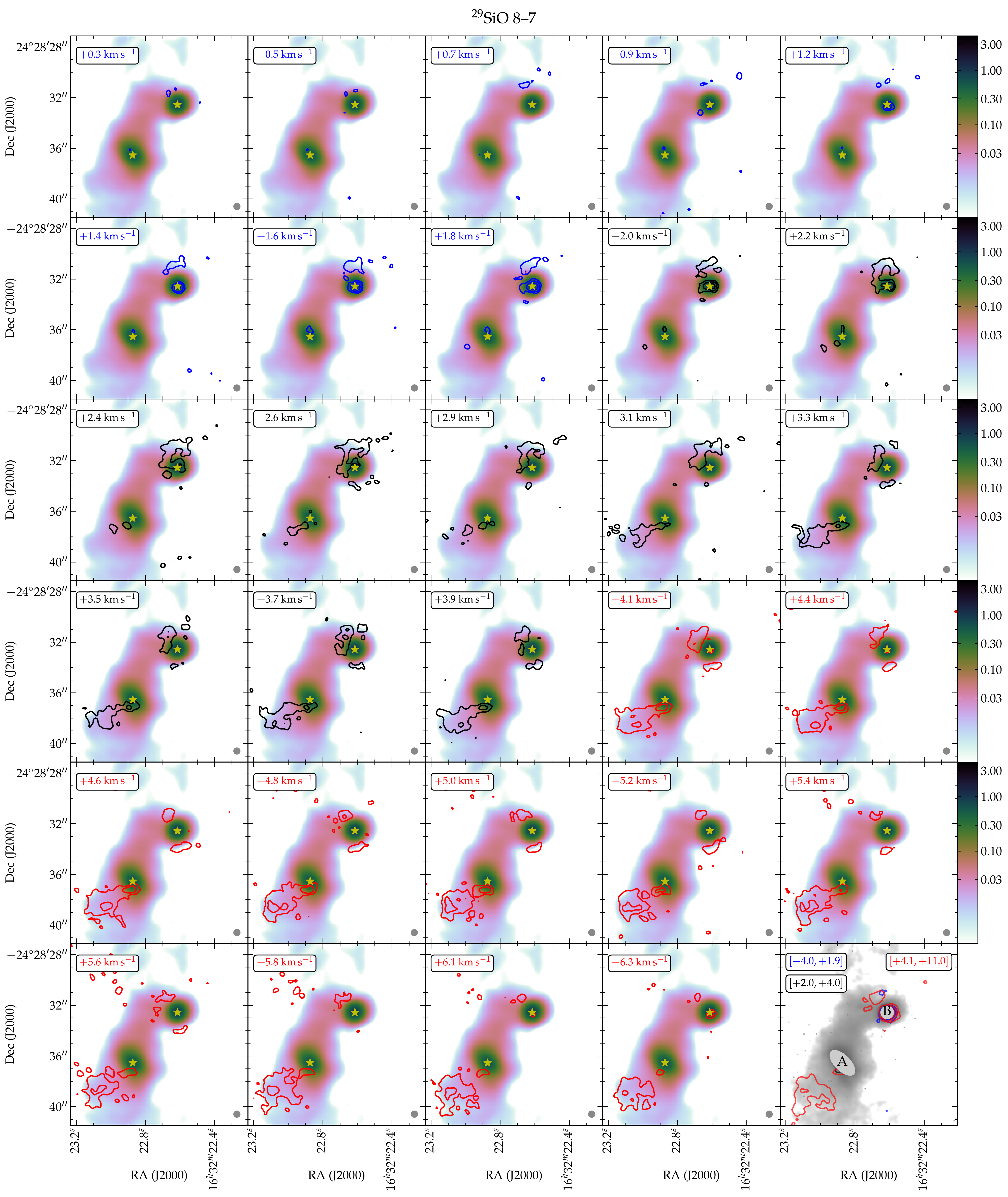}}
	\caption{
	$^{29}$SiO 8--7 channel maps. 
	As Fig.~\ref{fig:C17Ochannelmaps}. First positive (negative) contour level in individual channel maps: 0.02 ($-0.04$) \Jyperbeam; first contour levels in the bottom right integrated ranges panel: 0.07 \Jyperbeam\,\kms. 
	 }
	\label{fig:29SiOchannelmaps}
\end{figure*}

\begin{figure*}
	\resizebox{\hsize}{!}{\includegraphics[angle=0]{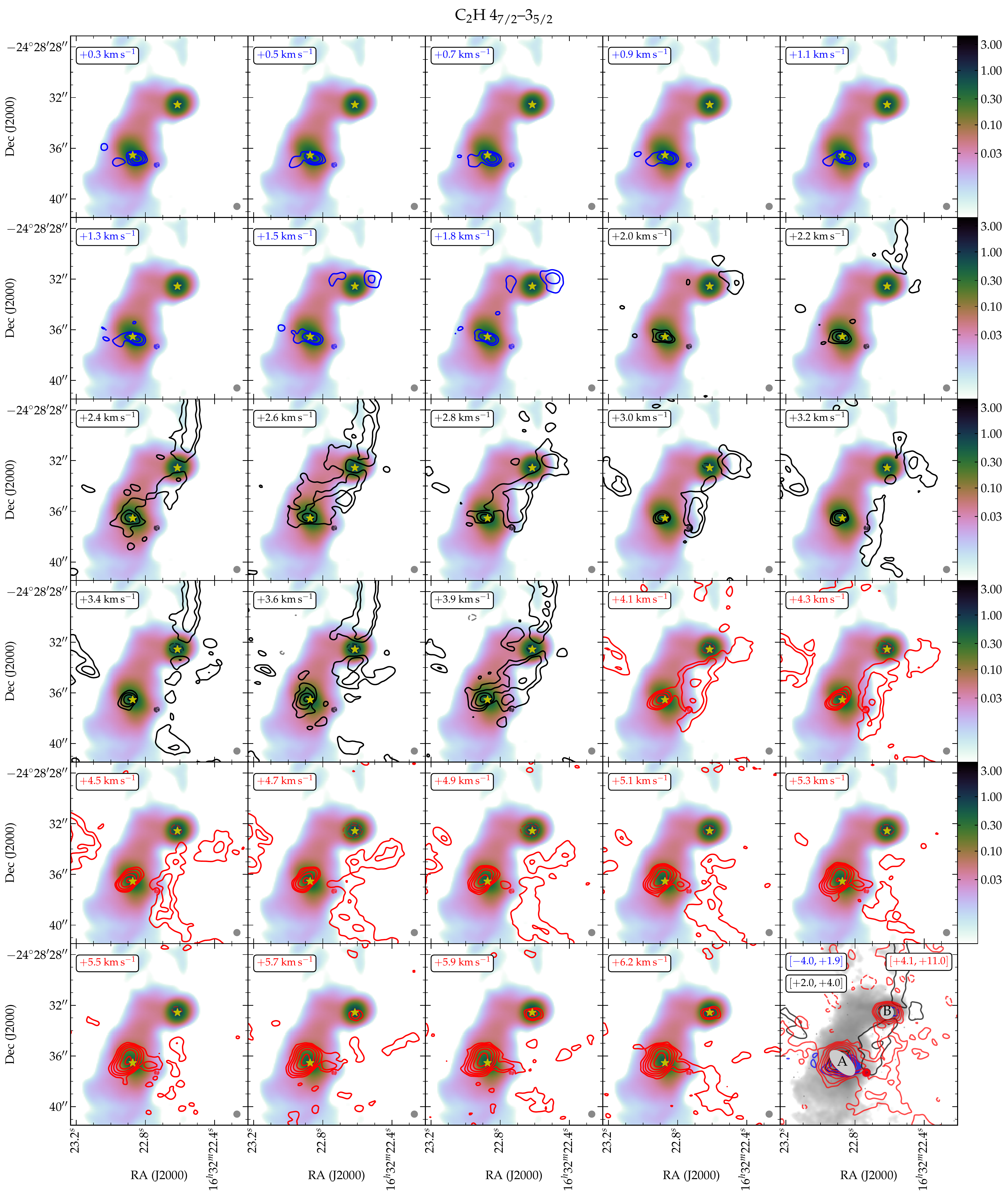}}
	\caption{
	\CCH\ channel maps. 
	As Fig.~\ref{fig:C17Ochannelmaps}. First positive (negative) contour level in individual channel maps: 0.03 ($-0.06$) \Jyperbeam; first contour levels in the bottom right integrated ranges panel: 0.07 \Jyperbeam\,\kms. 
	Note that emission in channels above +5~\kms\ is due to CH$_3$CN 19(3)--18(3) and not \CCH. 
	}
	\label{fig:CCHchannelmaps}
\end{figure*}

\clearpage

\section{Model maps without freeze-out}
\label{sec:nofreezeoutmaps}

In Fig.~\ref{fig:nofreezeoutmaps} we show, for completeness, the result of the radiative transfer model (Sect.~\ref{sec:bridgemodel}) with freeze-out effects discarded for o-\HHCO\ and \HthCN. The dashed contours, representing negative intensity values, indicate that optical depth of \HthCN~4--3 is high when all \HthCN\ of all model components is kept in the gas phase, as discussed in Sect.~\ref{sec:bridgemodel}. These `no freeze-out' calculations are included in this work with the aim of disentangling effects of excitation and those of freeze-out and sublimation. They are explicitly not meant to present an alternative physical scenario; freeze-out of molecules onto dust grain surfaces should always be taken into account in conditions representative of star-forming regions. 

\begin{figure}[!bth]
	\resizebox{\hsize}{!}{\includegraphics[angle=0]{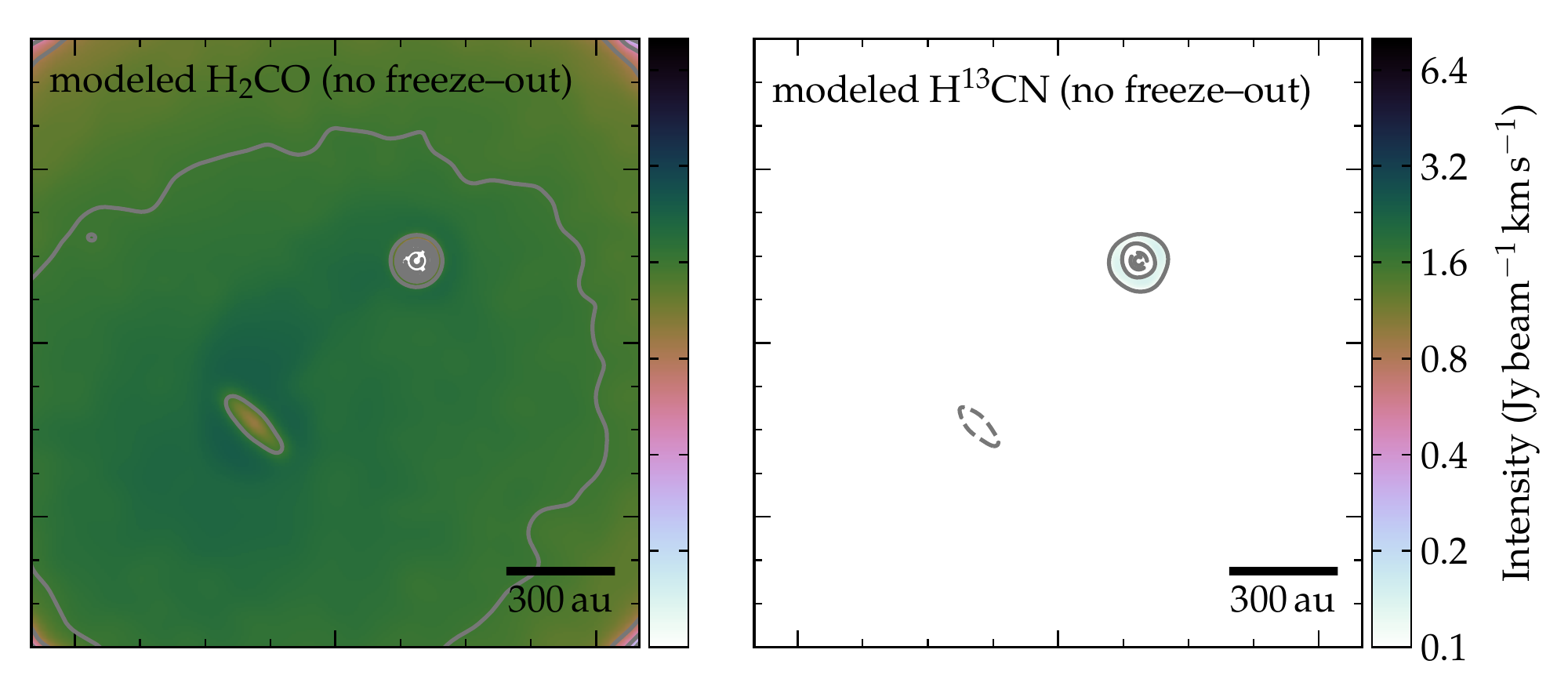}}
	\caption{
	Radiative transfer model maps (see Sect.~\ref{sec:bridgemodel}) of o-\HHCO~5$_{1,5}$--4$_{1,4}$ (left) and \HthCN~4--3 (right) with freeze-out effects ignored. These maps are counterparts of the more physically correct ones where freeze-out is implemented, in the middle right and bottom right panels of Fig.~\ref{fig:obs-vs-model}. Color scales and contours are identical to those used in Fig.~\ref{fig:obs-vs-model}, with the exception that negative intensity contours (dashed) are included in this figure. 
	}
	\label{fig:nofreezeoutmaps}
\end{figure}

\end{appendix}

\end{document}